\documentclass[3p]{elsarticle}

\usepackage{lineno,hyperref}
\usepackage{amsmath}
\usepackage{amssymb}
\usepackage{amsfonts}
\usepackage{graphicx}
\usepackage{graphicx,psfrag}
\usepackage{float}
\usepackage{overpic}
\usepackage{pgfplots}
\usetikzlibrary{patterns}
\usetikzlibrary{arrows,shapes,positioning}
\usepgfplotslibrary{groupplots}
\usepackage{wasysym}
\usepackage{mathtools}
\usepackage{rotating}
\usepackage{lmodern}
\usepackage{color}
\usepackage{stmaryrd}
\usepackage{afterpage}
\usepackage{tabularx}
\usepackage{multirow}
\usepackage{multicol}
\usepackage{caption}
\usepackage{subcaption}
\usepackage{framed}
\usepackage{bchart}
\usepackage{soul}
\usepackage{todonotes}
\usepackage{xspace}
\usepackage{soul}
\usepackage{pifont}
\usepackage{makecell}
\usepackage{siunitx}
\usepackage{placeins}
\usepackage[normalem]{ulem}
\usepackage{nomencl}

\usepackage{nomencl}
\makenomenclature
\renewcommand*\nompreamble{\begin{multicols}{2}}
\renewcommand*\nompostamble{\end{multicols}}

\newcommand{\B}[1]{\boldsymbol{#1}}
\newcommand{\mc}[1]{\mathcal{#1}}

\newcommand{\of}[1]{\!\left({#1}\right)}
\newcommand{\sbr}[1]{\left[{#1}\right]}

\newcommand{\dyad}{\otimes}
\newcommand{\commentout}[1]{}
\newcolumntype{L}[1]{>{\raggedright\arraybackslash}p{#1}}
\newcolumntype{C}[1]{>{\centering\arraybackslash}p{#1}}
\newcolumntype{R}[1]{>{\raggedleft\arraybackslash}p{#1}}
\newcommand{\Lemaitre}{\text{Lemaitre}\xspace}
\newcommand{\Bezier}{\text{B\'{e}zier}\xspace}
\newcommand{\Lame}{\text{Lame}\xspace}
\newcommand{\modelI}{\text{ECC}\xspace}
\newcommand{\modelII}{\text{LEM}\xspace}
\newcommand{\modelIa}{\text{ECC-a}\xspace}
\newcommand{\modelIb}{\text{ECC-i}\xspace}
\newcommand{\modelIIa}{\text{LEM-a}\xspace}
\newcommand{\modelIIb}{\text{LEM-i}\xspace}
\newcommand{\PEEQ}{$\ve^\text{p,eq}$\xspace}

\newcommand{\el}{^{\text{e}}}
\newcommand{\pl}{^{\text{p}}}

\newcommand{\mr}[1]{\mathrm{#1}}

\newcommand{\ve}{\varepsilon}
\newcommand{\mac}[1]{\left\langle{#1}\right\rangle}

\newcommand{\partDer}[2]{\dfrac{\partial #1}{\partial #2}}

\newcommand{\trace}[1]{\mr{tr} \left( #1 \right)}

\newcommand{\dev}[1]{\mr{dev} \left( #1 \right)}
\renewcommand{\exp}[1]{\mr{exp} \left( #1 \right)}

\newcommand{\abs}[1]{ \left| #1 \right|}

\definecolor{TRRRed}{RGB}{222,0,0}
\definecolor{TRRBlue}{RGB}{0, 104, 178}
\definecolor{TRROrange}{RGB}{250, 174, 40}
\definecolor{TRRGreen}{RGB}{135, 192, 33}
\definecolor{TRRPurple}{RGB}{148,0,211}
\definecolor{TRRCyan}{RGB}{0, 200, 255}
\definecolor{TRRGray}{RGB}{215, 215, 215}
\definecolor{TRRWhite}{RGB}{255, 255, 255}

\newcommand{\KF}[1]{{\color{black}{#1}}} 

\newtheorem{remark}{Remark}

\modulolinenumbers[5]

\journal{Engineering Fracture Mechanics}

\begin{document}

\begin{frontmatter}

\title{On the accuracy of isotropic damage descriptions for complex load paths -- Limits quantified by numerical optimization}

\author{K. Feike}
\author{P. Kurzeja}
\author{J. Mosler}
\author{K. Langenfeld\corref{CorrespondingAuthor}}

\cortext[CorrespondingAuthor]{Corresponding author. kai.langenfeld@tu-dortmund.de}

\address{TU Dortmund University, Institute of Mechanics, Leonhard-Euler-Str. 5, D-44227 Dortmund, Germany}

\begin{abstract}
	The stress triaxiality and the Lode angle parameter are two well established stress invariants for the characterization of damage evolution. At the same time, their limitations are known in principle as they are not sufficient, but remain to be quantified for a full damage description. This work contributes precisely by quantifying these restrictions through numerical optimization in terms of load paths and characteristic damage values. This is a key requirement for an accurate distinction between damage-mitigating and damage-prone scenarios. Such isotropic damage approximations -- among others -- are utilized in this work for damage predictions in a continuum damage mechanics framework. For that purpose, two well-established damage models are considered and analyzed for different complex load paths. Two models are used in order to avoid model-specific restrictions. For the different load paths, different isotropic damage indicators such as the stress triaxiality and the Lode angle parameter are controlled. The analyses show that well-established concepts such as fracture surfaces depending on the triple stress triaxiality, Lode angle parameter and equivalent plastic strain have to be taken with care, if complex paths are to be investigated. These include, e.g., load paths observed during metal forming applications with varying load directions or multiple stages.
\end{abstract}


\end{frontmatter}



\nomenclature[01]{\textit{\modelI}}{prototype model 1}
\nomenclature[02]{\textit{\modelII}}{prototype model 2}

\nomenclature[03]{$\bullet\el$/$\bullet\pl$}{elastic/plastic part}
\nomenclature[04]{$\bullet^\text{hyd}$}{hydrostatic part}
\nomenclature[05]{$\bullet^\text{dev}$}{deviatoric part}
\nomenclature[06]{$\bullet^\text{eq}$}{equivalent quantity}
\nomenclature[07]{$\bullet_{ij}$}{ij-th tensor component}
\nomenclature[08]{$\bullet_i$, $\B{N}_i$}{i-th eigenvalue and associated eigenvector}
\nomenclature[09]{$\bullet_+$/$\bullet_-$}{positive/negative part}
\nomenclature[10]{$\dot{\bullet}$}{time derivative}

\nomenclature[11]{$\eta$}{stress triaxiality}
\nomenclature[12]{$\bar{\theta}$}{Lode angle parameter}
\nomenclature[13]{$L$}{Lode parameter}

\nomenclature[14]{$I_i$}{i-th stress invariant}
\nomenclature[15]{$J_i$}{i-th deviatoric stress invariant}

\nomenclature[16]{$\psi$}{Helmholtz energy}
\nomenclature[17]{$\mathcal{G}$}{Gibb's energy}
\nomenclature[18]{$\mathcal{D}^\text{red}$}{reduced dissipation}

\nomenclature[19]{$\B{\varepsilon}$}{strain tensor}
\nomenclature[20]{${\B{\ve}}\pl$}{internal variable: plastic strain tensor}
\nomenclature[21]{$\B{\sigma}$}{stress tensor}

\nomenclature[22]{$\B{a}$}{internal variable: kinematic hardening}
\nomenclature[23]{$\B{\alpha}$}{back stress tensor}

\nomenclature[24]{$\B{k}$, $k$}{internal variable: isotropic hardening}
\nomenclature[25]{$\B{\kappa}$, $\kappa$}{drag stress (tensor)}

\nomenclature[26]{$\B{b}$, $b$}{internal variable: integrity (tensor)}
\nomenclature[27]{$\B{\beta}$, $\beta$}{energy release rate (tensor)}

\nomenclature[28]{$\B{D}$, $D$}{internal variable: damage (tensor)}
\nomenclature[29]{$\B{Y}$, $Y$}{energy release rate (tensor)}
\nomenclature[30]{$\B{H}$}{integrity tensor}

\nomenclature[31]{$\B{\tau}$}{relative stress tensor}

\nomenclature[32]{$g$}{plastic potential}
\nomenclature[33]{$\Phi$}{yield function}
\nomenclature[34]{$\Gamma_\bullet$}{non-associative parts of $g$}

\nomenclature[35]{$E$, $\nu$}{Young's modulus, Poisson's ratio}
\nomenclature[36]{$\lambda$, $\mu$}{Lam\'{e} parameter}

\nomenclature[37]{$\tau_y$}{initial yield stress}

\nomenclature[38]{$H_a$}{hardening modulus (kinematic)}
\nomenclature[39]{$B_a$}{yield stress saturation (kinematic)}

\nomenclature[40]{$H_i$}{hardening modulus (isotropic)}
\nomenclature[41]{$\Delta \tau$, $\kappa_u$}{yield stress saturation (isotropic)}

\nomenclature[42]{$C_\bullet$}{damage modulus}
\nomenclature[43]{$m$}{damage power law}

\nomenclature[44]{$\B{P}$, $\B{Q}$}{control points}
\nomenclature[45]{$t$}{normalized time}

\nomenclature[46]{$\mathbb{E}$}{4th-order elastic stiffness tensor}
\nomenclature[47]{$\mathbb{C}$}{4th-order elastic compliance tensor}

\nomenclature[48]{$\B{r}$}{direction vector}
\nomenclature[49]{$E_{\B{r}}$}{projected stiffness}
\nomenclature[50]{$C_{\B{r}}$}{projected compliance}

\nomenclature[51]{$\xi_{\mathbb{E}}$}{relative stiffness}
\nomenclature[52]{$\xi_{\mathbb{C}}$}{relative compliance}

\nomenclature[53]{$\delta_{ij}$}{Kronecker-Delta}
\nomenclature[54]{$\B{I}$}{2nd-order identity tensor}
\nomenclature[55]{$\mathbb{I}$}{4th-order symmetric identity tensor}

\nomenclature[56]{$\B{R}$}{rotation matrix}

\section{Introduction}\label{Sec:Introduction}
	The prediction of damage in materials is still a major challenge in many engineering disciplines since the underlying mechanisms are even nowadays not fully understood. As a result, large safety factors are mandatory to statistically ensure the functionality of structures and components. The modeling of the underlying material behavior is thus a key aspect for understanding and predicting the performance of these parts, cf. \cite{stephens_metal_2001,ma_computational_2015,tekkaya_damage_2020}.
	
	From a microscopic view, the mechanism of ductile damage can be related to the nucleation, the growth and finally the coalescence of voids induced by plastic deformation~\cite{KOPLIK1988835}. Depending on the material of interest, voids can show different origins and shapes, e.g., decohesion at hard particles. Since voids reduce the effective (load-carrying) area, they result in a reduced effective material stiffness~\cite{Sancho2016}, which ultimately also reduces the load-bearing capacity of the material. A natural modeling framework is thus the effective stress concept, cf.~\cite{lemaitre:hal-03609806, Gurson1977}. Within this concept, which is usually combined with the principle of strain equivalence, the true stresses are introduced. They are defined with respect to the effective cross sectional area due to void nucleation, growth and coalescence. An alternative modeling concept is based on effective strains, which can be interpreted as the decomposition of the total strains into elastic and damage-related parts~\cite{SimoJu1987, brunig_anisotropic_2003}. Finally, an effective damage-induced elastic stiffness can also be derived by means of the concept of strain energy equivalence~\cite{STEINMANN19981793, Menzel02, Ekh03}. In contrast to the concept of effective stresses and the concept of effective strains, the variational principle of strain energy equivalence naturally enforces physically relevant invariances. For instance, it naturally leads to a symmetric Cauchy stress even for anisotropic damage degradation (certainly, provided a Boltzmann continuum is considered). While these different principles lead to similar models for isotropic damage approximations, this is not the case for anisotropic damage models such as those considered in this paper.
	

	Turning the view to the macroscopic description, a common approach for damage characterization is based on the stress invariants triaxiality $\eta$ and Lode angle parameter $\bar{\theta}$
	\begin{align}\label{Eq:stress_invariants}
		\eta &= \dfrac{I_1}{\sqrt{3 \, J_2}} \, , \\
		\bar{\theta} &= 1 - \dfrac{2}{\pi} \, \arccos\of{L} \quad \text{with} \quad 
		L = \dfrac{J_3}{2} \, \left[ \dfrac{3}{J_2} \right]^{3/2} \; , \label{Eq:lode_angle}
	\end{align}
	and the equivalent plastic strain, cf.~\cite{wierzbicki_new_2005,BAI20081071},
	\begin{align}\label{Eq:eq_pl_strain}
		\ve^\mathrm{p,eq} &= \int \limits_{0}^t \sqrt{\dfrac{2}{3} \, \dot{\B{\ve}}\pl : \dot{\B{\ve}}\pl} \, \mathrm{d}\bar{t} \, \text{.}
	\end{align}
	\( I_1 \), \( I_2 \), \( I_3 \), \( J_2 \) and \( J_3 \) are thereby the usual invariants the stress tensor \( \B{\sigma} \) and given in terms of its eigenvalues \( \sigma_i \) as
	\begin{equation}
		I_1 = \sigma_1 + \sigma_2 + \sigma_3 \, , \quad I_2 = \sigma_1 \, \sigma_2 + \sigma_2\, \sigma_3 + \sigma_3 \, \sigma_1 \, , \quad I_3 = \sigma_1 \, \sigma_2 \, \sigma_3 \; ,
	\end{equation}
	\begin{equation}
		J_2 = \dfrac{1}{3} \, I_1^2 - I_2 \, , \quad J_3 = \dfrac{2}{27} \, I_1^3 - \dfrac{1}{3} \, I_1 \, I_2 + I_3 \, .
	\end{equation}
	From a mathematical point of view, a characterization by the aforementioned invariants corresponds to an isotropic approximation. 
	
	Early research showed that the stress triaxiality can be related to the physical process of void nucleation, growth and coalescence, cf.~\cite{McClintock1968, RICE1969}. Likewise, the Lode angle parameter was found to characterize damage evolution at low stress triaxialities, see~\cite{BAI20081071,brunig_micro-mechanical_2013}. More recently, the Lode angle parameter has been investigated for different fracture criteria in order to characterize the onset of damage~\cite{mattiello_lode_2021}. The entire triple of stress triaxiality, Lode angle parameter and equivalent plastic strain can be found in the context of so-called {\em fracture surfaces}~\cite{wierzbicki_new_2005,BAI20081071,anderson_failure_2017}. A fracture surface represents an iso-surface in the space of $\{\eta,\,\bar{\theta},\,\ve^\mathrm{p,eq}\}$, at which macroscopic failure is observed in experiments. This concept can also be adapted for damage models~\cite{basaran_stress_2011,andrade_incremental_2016,neukamm_lokalisierung_2018}. \KF{Several model modifications have been proposed to increase the prediction accuracy based on the stress triaxiality and Lode angle for more complex load paths~\cite{cao_lode-dependent_2014,zhang_enhanced_2021}. This is of utmost importance in several applications characterized by varying load paths, such as forming processes and service conditions, cf.~\cite{tekkaya_forming-induced_2017,meya_stress_2019,nick_numerical_2020}.}
	
	To investigate the effect of the stress state on ductile damage, loads at constant stress triaxiality have been analyzed in~\cite{kuna_three-dimensional_1996,pardoen_extended_2000,lin_performing_2006} with a focus on the micro scale. Further studies have extended the analyses to include or to additionally control the Lode (angle) parameter~\cite{zhang_numerical_2001,gao_modeling_2006,kiran_triaxiality_2014,cadet_ductile_2021}. Findings clearly indicate a dependency of the underlying microstructure (distribution of voids in a unit cell) on the macroscopic response~\cite{cadet_ductile_2021}. This indicates limits of common isotropic approaches for damage characterization.
	
	Among the variety of microscopic and macroscopic influences, stress triaxiality and Lode angle parameter hence contribute significantly to damage evolution~\cite{zhang_numerical_2001,gao_modeling_2006,kiran_triaxiality_2014,cadet_ductile_2021}. Despite their usefulness, though, it is well known that they are not sufficient -- for instance, due to additional influences such as anisotropic behavior. Moreover, the stress invariants stress triaxiality and Lode angle parameter often change in time for complex load paths. Therefore, such load paths -- as occurring in forming processes~\cite{hering_damage-induced_2020,gitschel_controlling_2023} -- require an extended description.
	
	Although isotropic damage approximations are well established, their limits remain to be fully explored and understood in terms of their impact on  engineering processes. The present work thus quantitatively and comprehensively analyzes the limits of simplified isotropic damage characterizations such as the triple triaxiality, Lode angle and equivalent plastic strain in Eqs.~\eqref{Eq:stress_invariants}--\eqref{Eq:eq_pl_strain} -- for the first time also quantitatively based on optimization algorithms.  It uses numerical models for this purpose, which we understand as one major scientific instrument that can guide selected experimental investigations. The numerical framework is calibrated with data from case-hardened steel 16MnCrS5. The present numerical approach provides the benefit of allowing broadband parameter studies. This numerical investigation shall thus systematically disclose the missing relationships and guide future experimental exploration and validation. The leading property of interest is ductile damage evolution in the context of continuum damage mechanics.
	
	The key highlights are:
	\begin{itemize}
		\item Computational framework to quantify limits of established isotropic damage description based on load path optimization
		\item In order to ensure model independent statements for real materials two different damage models (such as the Lemaitre-model~\cite{Lematre1992ACO}) are considered and calibrated on case hardened steel
		\item Computational optimization identified load paths with 13 \% deviation in final damage -- despite identical final stress triaxiality, Lode angle parameter and equivalent plastic strain  plastic strain
		\item Even an identical stress invariant history still allows for deviations up to 10 \%
	\end{itemize}
	
	In order to elaborate these highlights, two anisotropic damage models depending on the complete stress tensor are considered. Their prediction capabilities are then evaluated, for instance, by comparison of two different load paths with identical stress triaxialities and/or Lode angle parameters. Since such numerical experiments may also depend on the underlying constitutive model, two different modeling frameworks are adopted. While the first one is based on the principle of energy equivalence~\cite{STEINMANN19981793, Menzel02, Ekh03}, the second one is the by now classic \Lemaitre model~\cite{EngDamageMech05}.
	
	This work is structured as follows. Section 2 describes the two constitutive models based on continuum damage mechanics, which form the basis for the subsequent investigations. The numerical methodology for the discretization, control and optimization of load paths is presented in Section 3. Particularly, techniques for prescribing the stress-triaxiality and the Lode (angle) parameter are presented. The main findings are reported in Section 4 by numerical experiments that quantify the limits of different isotropic damage characterizations. Section 5 concludes with implications for the requirement of a more general damage characterization.


\section{Continuum damage mechanics -- two prototype models}\label{Sec:Fundamentals}

	Two constitutive models are briefly presented for the prediction of ductile damage, both as isotropic and anisotropic formulations. This allows a comprehensive load path analysis that is not bound to the characteristics of a single model. The first model, \textit{\modelI}, is a characteristic prototype of the effective configuration concept (on the basis of the principle of strain energy equivalence). The second model, \Lemaitre model \textit{\modelII}, is a characteristic prototype of the effective stress concept (combined with the principle of strain equivalence). All models have been calibrated with the same experimental data and form the basis for the subsequent numerical investigations.

	\subsection{ECC-Model -- Effective Configuration Concept}\label{Ssec:Model_ekh03}
		\newcommand{\ieps}{\B{b} : \B{\ve}\el_+ + \B{I}:\B{\ve}\el_-}
		\newcommand{\iieps}{\B{b} : \left[ \B{\ve}\el_+ \cdot \B{b} \cdot \B{\varepsilon}\el_+\right] + \B{\ve}\el_-:\B{\ve}\el_- }
		\newcommand{\ikk}{\B{b} : \B{k}}
		\newcommand{\iia}{\B{b} : \left[ \B{a} \cdot \B{b} \cdot \B{a}  \right]}
		\newcommand{\iiepsp}{\B{b} : \left[ \B{\ve}\pl \cdot \B{b} \cdot \B{\ve}\pl \right]}
		The first model is named after the effective configuration concept and is based on the principle of strain energy equivalence, cf.~\cite{STEINMANN19981793}. The constitutive relations are adopted from the works in \cite{Menzel02, Ekh03}.
		The Helmholtz energy $\psi$ is additively decomposed into an elastic part ($\psi\el$) and a plastic part ($\psi\pl$) according to
		\begin{align}
			\psi &= \psi\el(\B{\ve}, \, \B{\ve}\pl, \, \B{b}) + \psi\pl( \B{a}, \, \B{k}, \, \B{b}, \, \B{\ve}\pl) \, \text{,}
			\\
			\psi\el &= \dfrac{\lambda}{2} \, \left[\ieps\right]^2 + \mu \, \left[\iieps\right] \, \text{,}
			\\
			\psi\pl &= \dfrac{H_i}{2} \, \left[\ikk \right]^2 + \dfrac{H_a}{2} \, \iia \text{.} 
		\end{align}
		The strain tensor $\B{\ve}$ is also additively decomposed into an elastic and a plastic part, i.e., $\B{\ve} = \B{\ve}\el + \B{\ve}\pl$, cf.~\cite{GreenNaghdi1965}.
		The model captures isotropic as well as kinematic hardening due to the state variables $\B{k}$ and $\B{a}$, respectively.
		Here, isotropic hardening is captured by a tensorial internal variable, following the principle of strain energy equivalence \cite{Ekh03}. Both the elastic and the plastic energy contribution are affected by the damage evolution due to integrity tensor $\B{b}$. The integrity tensor can be interpreted as an inverse damage measure, i.e., $\B{b} = \B{I}$ represents the virgin material, while at least one of the eigenvalues of $\B{b}$ converges to zero for completely damaged material points. By choosing $\B{b} = \B{I}$, Hooke's law is recovered. Accordingly, $\lambda$ and $\mu$ are the \Lame parameters. Furthermore, $H_i$ is the isotropic hardening modulus and $H_a$ is the kinematic hardening modulus.
		\\
		A crucial contribution to the material model is the micro crack-closure-reopening (MCR)-effect, which only allows damage evolution under tensile modes. Within the present model, the MCR effect is incorporated through an additive decomposition of the elastic strain tensor into its tensile and its compression modes according to
		\begin{align}\label{eq:mcr_decomp}
			\B{\ve}\el = \B{\ve}\el_+ + \B{\ve}\el_- \text{,} \quad \B{\ve}\el_+ &= \sum_{i=1}^{3} \mathcal{H}\of{\ve\el_i} \B{N}_i \otimes \B{N}_i.
		\end{align}
		Here, $\ve\el_i$ are the eigenvalues, $\B{N}_i$ are the associated eigenvectors of the elastic strain tensor and $\mathcal{H}$ is the Heaviside function.
		Within the numerical implementation, the discontinuous Heaviside function has been approximated in line with~\cite{Ekh03} by setting numerical parameters \( g_0 = 0 \), \( x_0 = 0 \) and \( x_R = 10^{-6} \). It bears emphasis that the considered anisotropic MCR model allows to capture damage evolution, even if the stress triaxiality is negative. A straightforward calculation of the thermodynamic driving forces leads to
		\begin{align}
			\B{\sigma}\phantom{\el} &=\phantom{-}\partDer{\psi\phantom{\el}}{\B{\ve}} = \lambda \, \left[\ieps\right] \, \left[ \B{b} : 	 \partDer{\B{\ve}\el_+}{\B{\ve}\el} + \B{I} : \partDer{\B{\ve}\el_-}{\B{\ve}\el}\right] + 2\,\mu\, \left[ \left[ \B{b} \cdot \B{\ve}\el_+ \cdot \B{b} \right] : \partDer{\B{\ve}\el_+}{\B{\ve}\el} + \B{\ve}\el_- : \partDer{\B{\ve}\el_-}{\B{\ve}\el} \right] \, \text{,}
			\\
			\B{\alpha}\phantom{\el} &= -\partDer{\psi\phantom{\el}}{\B{a}} = - H_a\, \B{b} \cdot \B{a} \cdot \B{b}\, \text{,}
			\\
			\B{\kappa}\phantom{\el} &= -\partDer{\psi\phantom{\el}}{\B{k}} = - H_i \, \left[ \B{b} : \B{k} \right]\, \B{b}\, \text{,}
			\\
			\B{\beta}\el &= -\partDer{\psi\el}{\B{b}} = - \lambda \, \left[ \B{b} : \B{\ve}\el_+ + \B{I} : \B{\ve}\el_- \right] \B{\ve}\el_+ - 2 \,  \mu \, 	\B{\ve}\el_+ \cdot \B{b} \cdot \B{\ve}\el_+ \, \text{,}
			\\
			\B{\beta}\pl &= -\partDer{\psi\pl}{\B{b}} = - H_a \, \B{a} \cdot \B{b} \cdot \B{a} - H_i \, \left[ \B{b} : \B{k} \right] \,  \B{k} \, 	\text{,}
			\\
			\B{\beta}\phantom{\el} &= \B{\beta}\el + \B{\beta}\pl \, \text{,}
		\end{align}
		where $\B{\sigma}$ is the stress tensor, $\B{\alpha}$ is the back stress tensor, $\B{\kappa}$ is the drag stress tensor and $\B{\beta}$ the energy release rate. Using the thermodynamic driving forces, the dissipation inequality reads in its reduced form
		\begin{align}
			\mc{D}^\text{red} &= \B{\sigma} : \dot{\B{\ve}} - \dot{\psi} = \B{\sigma} : \dot{\B{\varepsilon}}\pl + \B{\alpha} : \dot{\B{a}} +  \B{\kappa} : \dot{\B{k}} + \B{\beta} : \dot{\B{b}} \geq 0.
		\end{align}
		By following the principle of Generalized Standard Materials \cite{Halphen1975SurLM}, a plastic potential is introduced for the definition of the evolution equations. The dissipation inequality is known to be automatically fulfilled if that potential is convex, non-negative and contains the origin. The plastic potential is specified as
		\begin{align}
			g &= \Phi(\B{\sigma}, \, \B{\alpha}, \, \B{\kappa, \, \B{b}}) + \Gamma_\alpha(\B{\alpha}, \, \B{b}) + \Gamma_\beta(\B{\beta}\el, \,  \B{b})\textbf{,}  \label{Ekh:PlastPot}
		\end{align}
		and consists of the yield function $\Phi$ and non-associated parts $\Gamma_{\!\alpha}$ and $\Gamma_\beta$.
		The yield function is chosen as
		\begin{alignat}{3}
			& &&\Phi &&= \sqrt{\bar{\tau}} - \tau_y - \frac{1}{3} \, \B{b}^{-1} : \B{\kappa} - \Delta \tau \, \left[ 1 - \exp{-\frac{\left\vert  \B{b}^{-1}  : \B{\kappa} \right\vert}{\kappa_u}} \right] \, \text{,} \label{Eq:Ekh_yield}
			\\
			& &&\bar{\tau} &&= \dfrac{3}{2} \, \B{b}^{-1} : \left[ \B{\tau} \cdot \B{b}^{-1} \cdot \B{\tau} \right] - \dfrac{1}{2} \, \left[  \B{b}^{-1} :  \B{\tau} \right]^2 \quad \text{with } \B{\tau} = \B{\sigma} - \B{\alpha} \, \text{,}
		\end{alignat}
		where the equivalent stress $\sqrt{\bar{\tau}}$ is of von Mises-type and depends on the relative stress tensor $\B{\tau}$.
		According to Eq.~\eqref{Eq:Ekh_yield}, the yield function captures linear and exponential isotropic hardening. The respective model parameters are denoted as $\tau_y$, $\Delta \tau$ and $\kappa_u$.
		The non-associative parts of potential~\eqref{Ekh:PlastPot} are given as
		\begin{align}
			\Gamma_\alpha &= \dfrac{B_a}{2 \, H_a} \, \B{b}^{-1} : \left[ \B{\alpha} \cdot \B{b}^{-1} \cdot \B{\alpha} \right] \, \text{,}
			\\
			\Gamma_\beta &= \dfrac{C_i}{2} \, \left[\B{b}^m : \B{\beta}\el\right]^2 + \dfrac{C_a}{2} \, \B{b}^m : \left[ \B{\beta}\el \cdot  \B{b}^m  \cdot \B{\beta}\el \right]\, \text{,} \label{Eq:Ekh03:Gamma_beta}
		\end{align}
		and extend the model to non-linear kinematic hardening of Armstrong-Frederick type by means of model parameter $B_a$, cf.~\cite{frederick_mathematical_2007}.
		The damage evolution is controlled by potential $\Gamma_\beta$ and consists of an isotropic part (damage modulus $C_i$) and an anisotropic part (damage modulus $C_a$).
		The damage exponent $m$ allows for greater flexibility when calibrating the model to experiments. Furthermore, potential $\Gamma_\beta$ depends only on the elastic part of the energy release rate $\B{\beta}$. This novel modification of the original model ensures that damage evolves only if tension occurs in a certain direction (tension in the sense of the elastic strain tensor). The evolution equations follow as gradients of plastic potential~\eqref{Ekh:PlastPot} -- in line with the principle of Generalized Standard Materials -- and read
		\begin{align}
			\dot{\B{\varepsilon}}\pl = \dot{\lambda} \, \partDer{g}{\B{\sigma}}, \quad \dot{\B{a}} = \dot{\lambda} \, \partDer{g}{\B{\alpha}}, \quad  \dot{\B{k}} = \dot{\lambda} \, \partDer{g}{\B{\kappa}}, \quad \dot{\B{b}} = \dot{\lambda} \, \partDer{g}{\B{\beta}} = \dot{\lambda} \, \partDer{g}{\B{\beta}\el} \, \text{,}
		\end{align}
		which concludes the constitutive model.
	
		\paragraph*{Isotropic model}\label{Ssec:Model_ekh03i}
			The isotropic version of the anisotropic \modelI model follows straightforward by setting anisotropic damage modulus \(C_a = 0\). This model will also be employed within the numerical experiments.

	\subsection{LEM-Model -- Effective Stress Concept (\Lemaitre model)}\label{Ssec:Model_Lad}
		\newcommand{\isigp}{\left\langle\sigma^\mr{hyd}\right\rangle}
		\newcommand{\isign}{\left\langle - \sigma^\mr{hyd} \right\rangle}
		\newcommand{\iisigp}{\B{H}:\left[\B{\sigma}_+^\mr{dev} \cdot \B{H} \cdot \B\sigma_+^\mr{dev}\right]}
		\newcommand{\iisign}{\B\sigma_-^\mr{dev}:\B\sigma_-^\mr{dev}}
		\renewcommand{\iia}{\B{a} : \B{a}}
		\newcommand{\iialp}{\B{\alpha} : \B{\alpha}}
		The second model is based on the effective stress concept with equivalent strains and is adopted from~\cite{EngDamageMech05}. Gibbs energy $\mathcal{G}$ is specified as
		\begin{align}
			\mathcal{G} &= \mathcal{G}\el + \B{\sigma} : \B{\ve}\pl - \psi\pl \, \text{,} \label{Eq:Lad:Gibbs}
			\\
			\mathcal{G}\el &= \frac{1+2\, \nu}{2\, E} \, \left[ \iisigp + \iisign \right] + \frac{1-2\, \nu}{2\, E} \, \left[ \frac{\isigp^2}{1- D^\mr{hyd}} + \isign^2 \right] \, \text{,} \label{Eq:Lad:Gibbs_el}
			\\
			\psi\pl &= \dfrac{H_a}{2} \, \iia + \Delta \tau \, \left[k + \kappa_u\, \exp{-\dfrac{k}{\kappa_u}}\right] \label{Eq:Lad:psi_pl} \, \text{,}
		\end{align}
		and depends on model parameters Young's modulus $E$, Poisson's ratio $\nu$, the isotropic hardening parameters $\Delta \tau$ and $c$, and the kinematic hardening modulus $K$. The additive decomposition of the total strains into elastic and plastic parts is again considered as $\B{\ve} = \B{\ve}\el + \B{\ve}\pl$. Furthermore, the stress tensor is decomposed into a deviatoric ($\B{\sigma}^\mr{dev}$) and a hydrostatic part ($\sigma^\mr{hyd} \, \B{I}$) and subsequently into their respective positive and negative parts in order to capture the MCR effect. Strain-like variables $\B{a}$ and $p$ are suitable for kinematic and isotropic hardening, respectively. The damage evolution is captured by means of second-order tensor $\B{D}$, which enters energy~\eqref{Eq:Lad:Gibbs_el} in terms of variables
		\begin{align*}
			D^\mr{hyd} &\coloneqq \frac{1}{3} \, \trace{\B{D}}\text{,} \qquad \qquad \B{H} \coloneqq \left[\B{I} - \B{D}\right]^{-\frac{1}{2}} \,  \text{.}
		\end{align*}
		While undamaged material is represented by $\B{D}\!=\!\B{0}$, at least one eigenvalue of $\B{D}$ converges to one for completely damaged material points.
		
		The thermodynamic driving forces follow as the derivatives of $\mathcal{G}$ and read 
		\begin{align}
			\phantom{-} \B{\ve}\el &= \phantom{-} \partDer{\mathcal{G}}{\B{\sigma}} = \frac{1+\nu}{E} \, \left[ \dev{\B{H} \cdot \B{\sigma}_+^\mr{dev}  \cdot \B{H}} + \B{\sigma}_-^\mr{dev} \right] + \frac{1-2\, \nu}{E} \, \left[ \frac{\isigp}{1- D^\mr{hyd}} - \isign \right] \, \B{I} \, \text{,}\label{Lad:epse}
			\\
			\phantom{-} \B{\alpha} &=   - \partDer{\mathcal{G}}{\B{a}} = \frac{2}{3} \, 
			H_a \, \B{a} \, \text{,}
			\\
			\phantom{-} \kappa &= - \partDer{\mathcal{G}}{k} = \Delta \tau \, \left[1 - \exp{-\dfrac{k}{\kappa_u}}\right] \, \text{,}
			\\
			\phantom{-} \B{Y}  &=  - \partDer{\mathcal{G}}{\B{D}} = -\frac{1+2\,\nu}{E} \, \left[\B{\sigma}^\mr{dev}_+ \cdot \B{H} \cdot  \B{\sigma}^\mr{dev}_+\right] : \partDer{\B{H}}{\B{D}} - \frac{1-2\,\nu}{6\,E} \, \frac{\left\langle \sigma^\mr{hyd}\right\rangle^2}{\left[1 - D^\mr{hyd}\right]^2} \, \B{I} \, \text{.}
		\end{align}
		The transformation of Gibbs energy $\mathcal{G}$ into Helmholtz energy $\psi$ reads
		\begin{align}
			\psi &= \B{\sigma} : \B{\ve} - \mathcal{G} = \B{\sigma}:\B{\ve} - \mathcal{G}\el - \B{\sigma}:\B{\ve}\pl + \psi\pl \, \text{,}
		\end{align}
		and allows to derive the reduced dissipation inequality as
		\begin{align}
			\mathcal{D}^\mathrm{red} &= \B{\sigma}:\dot{\B{\ve}} - \dot{\psi} = \B{\sigma}:\dot{\B{\ve}}\pl - \B{Y} : \dot{\B{D}} -  \B{\alpha}:\dot{\B{a}} - \kappa \, \dot{k} \geq 0  \, \text{.}
		\end{align}
		Following the principle of Generalized Standard Materials once again, plastic potential $g$ is specified. Analogously to the previous model, $g$ is assumed to be of type
		\begin{align}
			g &= \Phi\of{\B{\sigma},\B{\alpha},\kappa,\B{D}} + \Gamma_{\!\alpha}\of{\B{\alpha}}\label{eq:lad:plasticpotential} \, \text{,}
		\end{align}
		with von Mises-type yield function
		\begin{align}
			\Phi &= \sqrt{\bar{\tau}} - \tau_y - \kappa \quad \text{with} \,\, \bar{\tau} = \frac{3}{2} \, \B{\tau}:\B{\tau} \text{,} \quad \B{\tau} =  \B{H} \cdot \B{\sigma}^\mr{dev} \cdot \B{H} - \B{\alpha}\, \text{,} \label{Eq:Lad:yield}
		\end{align}
		and initial yield limit \( \tau_y \). The non-associative part of potential~\eqref{eq:lad:plasticpotential} reads
		\begin{align}
			\Gamma_{\!\alpha} &= \frac{3 \, B_a}{4 \, H_a} \, \iialp \, \text{,}
		\end{align}
		and extends the model to non-linear kinematic hardening due to model parameter $B_a$. The evolution equations then result in
		\begin{align}
			\dot{\B{\ve}}\pl &=  \dot{\lambda} \, \partDer{g}{\B{\sigma}} \, \text{,} \qquad
			\dot{\B{a}} = -\dot{\lambda} \, \partDer{g}{\B{\alpha}} \, \text{,} \qquad
			\dot{k} = -\dot{\lambda} \, \partDer{g}{\kappa} = \dot{\lambda} \, \text{.}
		\end{align}
		Following~\cite{EngDamageMech05}, the evolution equation associated with damage tensor $\B{D}$ does not follow from potential \( g \), but is explicitly specified as 
		\begin{align}
			\dot{\B{D}} &= \left[C \, \bar{Y}\right]^m \, \left[ \dot{\B{\ve}}\pl_+ - \dot{\B{\ve}}\pl_- \right] \, \text{,}  \label{Eq:Lad:evolution}
			\\
			\text{with } \bar{Y} &= \frac{1 + \nu}{2\,E} \, \trace{ \left[ \B{H} \cdot \B{\sigma}^\mr{dev}_+ \cdot \B{H} \right]^2 } + \frac{3 -  6\,\nu}{2\,E} \, \frac{\mac{\sigma^\mr{hyd}}^2}{\left[1 - D^\mr{hyd}\right]^2}\, \text{,} \label{eq:Ybar}
		\end{align}
		and depends on model parameters $C$ and $m$. This evolution is formulated in terms of a scalar-valued energy release rate $\bar{Y}$ and the additive decomposition of the plastic strain rate tensor $\dot{\B{\ve}}\pl = \dot{\B{\ve}}\pl_+ + \dot{\B{\ve}}\pl_-$, cf.~Eq.~\eqref{eq:mcr_decomp}. In line with the \modelI model, the Lemaitre model can predict damage evolution even for negative stress triaxialities. This is again related to the additive decomposition of the deviatoric stress tensor into a positive and a negative part.
		
		\paragraph*{Isotropic model}\label{Ssec:Model_Lid}
			The isotropic version of the anisotropic \modelII model results through the modification of evolution Eq.~\eqref{Eq:Lad:evolution} into a spherical counterpart. Here, the choice 
			\begin{align}
				\dot{\B{D}} &= \left[\hat{C} \, \bar{Y}\right]^{\hat{m}} \, \dot{\ve}^\mr{p,eq} \, \B{I} \, \text{,}
			\end{align}
			is made since it leads to similar results as the underlying anisotropic model when considering a uniaxial tensile test.
		
	\subsection{Similarities and differences between the prototype damage models}\label{Ssec:Const_differences}
	The similarities of both models are the mechanisms they capture, i.e., isotropic and kinematic hardening as well as an anisotropic ductile damage evolution which is coupled to the evolution of plasticity. Furthermore, both models have been calibrated on the same experimental data. Despite that, the two models differ in the manner in which the mechanisms interact. 
    A key distinction lies in the influence of the damage metric (\( \B{b} \) and \( \B{D} \), respectively) on the Helmholtz free energy. In the \modelI model, the damage metric scales the entire Helmholtz energy, including the plastic part of the Helmholtz energy, whereas in the \modelII model, it only affects the elastic contribution. Moreover, the polynomial degree of the damage metric is different. Furthermore, the MCR effect is realized differently in the two models: via a spectral decomposition of the elastic strains in the \modelI model and via a spectral decomposition of the stresses in the \modelII model. The thermodynamic driving force for damage evolution also varies in terms of the evolution direction. While it is the elastic part of the energy release rate (direction of elastic strains) in the \modelI model, it is based on the plastic strains in the \modelII model.
    \KF{Focusing on isotropic damage evolution ($\B{b}=\bar{b}\,\B{I}$, $\B{D}=\bar{D}\,\B{I}$) and ignoring the MCR effect, the driving force of both models can be written as
 	\begin{align}
 		\beta\el_\mr{iso, red} &= \frac{J_2}{E \, \bar{b}^3} \, \left[1 + \nu - \left[1 - 2\,\nu\right]\,\eta^2\right] \, \text{,}\\
 		\bar{Y}_\mr{iso, red} &= \frac{J_2}{2\, E\, \left[1-\bar{D}\right]^2} \, \left[ 2 \, \left[1 + \nu \right] + \left[1-2\,\nu\right] \eta^2 \right] \, \text{.}
 	\end{align}
 	Accordingly, both models show a similar structure, although they are not identical. In both models, damage evolution is governed by the same polynomial order of $J_2$ and stress triaxiality $\eta$. Certainly, the practical relevance of this analogy is limited, since anisotropic effects and the MCR effect are indeed required in many cases.
	}
		
	\subsection{Model calibration with case-hardened steel 16MnCrS5}\label{Ssec:PI}
		\setlength{\parindent}{0pt}
		
\begin{figure}[ht]%
			\centering
			\hfill
			\begin{subfigure}[ht]{0.3\textwidth}
				\psfrag{tensx}[c][c]{\scalebox{0.7}{Tensile strain $\ve_{11}$ [-]}}
				\psfrag{tensy}[c][c]{\scalebox{0.7}{Tensile stress $\sigma_{11}$ [\si{\mega\pascal}]}}
				\psfrag{exp}[l][l]{\scalebox{0.7}{Experiment}}
				\psfrag{modelekh}[l][l]{\scalebox{0.7}{\modelIa model}}
				\psfrag{modeleki}[l][l]{\scalebox{0.7}{\modelIb model}}
				\includegraphics[width=0.8\textwidth]{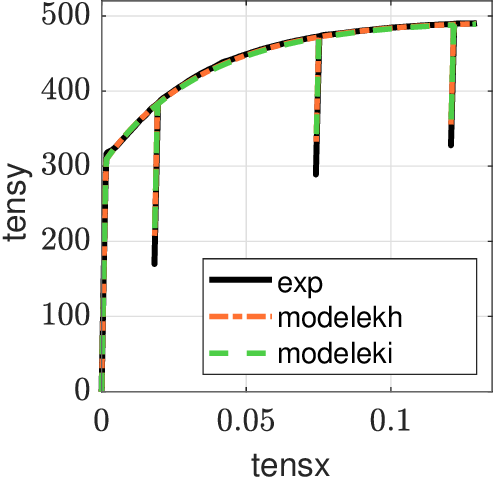}
				\caption{Tensile test: Isotropic and anisotropic \modelI model.}
			\end{subfigure}
			\hfill
			\hspace{0.2cm}
			\begin{subfigure}[ht]{0.3\textwidth}
				\psfrag{tensx}[c][c]{\scalebox{0.7}{Tensile strain $\ve_{11}$ [-]}}
				\psfrag{tensy}[c][c]{\scalebox{0.7}{Tensile stress $\sigma_{11}$ [\si{\mega\pascal}]}}
				\psfrag{exp}[l][l]{\scalebox{0.7}{Experiment}}
				\psfrag{modellad}[l][l]{\scalebox{0.7}{\modelIIa model}}
				\psfrag{modellid}[l][l]{\scalebox{0.7}{\modelIIb model}}
				\includegraphics[width=0.8\textwidth]{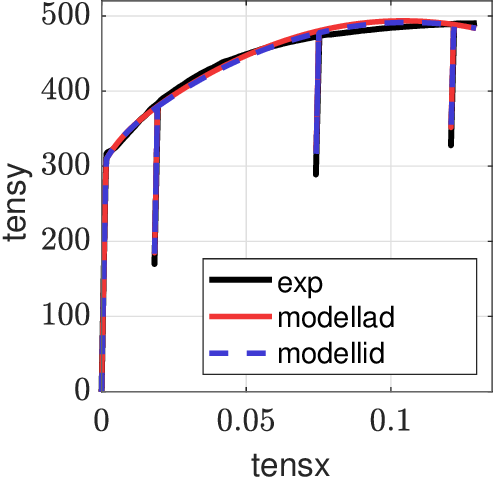}
				\caption{Tensile test: Isotropic and anisotropic \modelII model.}
			\end{subfigure}
			\hfill
			\begin{subfigure}[ht]{0.3\textwidth}
				\centering
				\psfrag{shearx}[c][c]{\scalebox{0.7}{Shear strain $\ve_{12}$ [-]}}
				\psfrag{sheary}[c][c]{\scalebox{0.7}{Shear stress $\sigma_{12}$ [\si{\mega\pascal}]}}
				\psfrag{modelekh}[l][l]{\scalebox{0.7}{\modelIa model}}
				\psfrag{modeleki}[l][l]{\scalebox{0.7}{\modelIb model}}
				\psfrag{modellad}[l][l]{\scalebox{0.7}{\modelIIa model}}
				\psfrag{modellid}[l][l]{\scalebox{0.7}{\modelIIb model}}
				\includegraphics[width=0.8\textwidth]{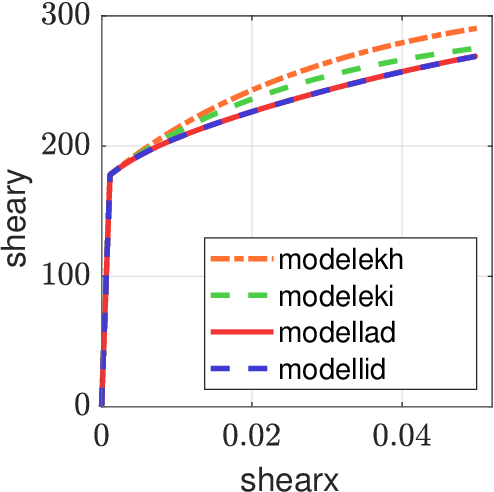}
				\caption{\KF{Numerical} shear test: All models.\\[4pt]\phantom{menace}}
			\end{subfigure}
			\hfill
			\caption{Experiment vs. simulation in calibration: \KF{(a) and (b)} Uniaxial tensile test with sequential unloading. \KF{(c) Additional numerical shear test comparing the model predictions.}}\label{Fig:Fits}
		\end{figure}

		All four models -- the isotropic and the anisotropic \modelI model (\modelIb and \modelIa) as well as the isotropic and the anisotropic \modelII model (\modelIIb and \modelIIa) -- are calibrated to experimental data from~\cite{LAN20} for comparability. This involves tensile experiments of case-hardened steel 16MnCrS5 up to strain amplitudes of $\ve_{11}\!\approx0.13$, which represents a typical material used in bulk forming such as forward extrusion. The strains remain spatially homogeneous in this deformation range and necking does not occur. All four models were calibrated by means of a standard least-squares approach. All models share the same elastic parameters $\lambda$ and $\mu$ and the same initial yield stress $\tau_\mathrm{y}$. Isotropic and anisotropic variants of the same model also share the same material parameters for plasticity. A visualization of the calibrations is shown in Fig.~\ref{Fig:Fits} and the final list of model parameters is given in Tabs.~\ref{Tab:Fundamentals:ModelParametersEkhAniso} and \ref{Tab:Fundamentals:ModelParametersLemAniso}.
		\begin{table}[ht]%
			\def\arraystretch{0.7}
			\centering
			\begin{tabular}{| L{5.3cm} | C{1cm} | C{1.9cm} | C{1cm} |}
				\hline
				Parameter name & Symbol & Value & Unit \\
				\Xhline{1.5pt}
				First \Lame parameter & $\lambda$ & $118870$ & \si{\mega\pascal}\\
				Second \Lame parameter & $\mu$ & $79249$ & \si{\mega\pascal}\\
				Yield stress & $\tau_y$ & $308.260$ & \si{\mega\pascal}\\
				Kinematic hardening modulus & $H_a$ & $7728.863$ & \si{\mega\pascal}\\
				Kinematic yield stress saturation & $B_a$ & $38.218$ & -\\
				Isotropic hardening modulus & $H_i$ & $1.829 \cdot 10^{-4}$ & \si{\mega\pascal}\\
				Isotropic hardening increment & $\Delta \tau$ & $2.261 \cdot 10^{-2}$ & \si{\mega\pascal}\\
				Isotropic hardening parameter & $\kappa_u$ & $0.159$ & \si{\mega\pascal}\\
				\hline
				\multicolumn{4}{| c |}{Anisotropic (\modelIa model)}\\
				\hline
				Isotropic damage modulus & $C_i$ & $0.0$ & \si{\mega\pascal^{-1}}\\
				Anisotropic damage modulus & $C_a$ & $14.503$ & \si{\mega\pascal^{-1}}\\
				Damage exponent & $m$ & $11.217$ & -\\
				\hline
				\multicolumn{4}{| c |}{Isotropic (\modelIb model)}\\
				\hline
				Isotropic damage modulus & $\hat{C}_i$ & $14.408$ & \si{\mega\pascal^{-1}}\\
				Anisotropic damage modulus & $\hat{C}_a$ & $0.0$ & \si{\mega\pascal^{-1}}\\
				Damage exponent & $\hat{m}$ & $11.373$ & -\\
				\hline
			\end{tabular}
			\caption{Model parameters of the \modelI model for case-hardenend steel 16MnCrS5. Parameters associated with the isotropic damage model are indicated with a superposed hat $\hat\bullet$.}\label{Tab:Fundamentals:ModelParametersEkhAniso}
		\end{table}%
		\begin{table}[ht]%
			\def\arraystretch{0.7}
			\centering
			\begin{tabular}{| L{5.3cm} | C{1cm} | C{1.9cm} | C{1cm} |}
				\hline
				Parameter name & Symbol & Value & Unit \\
				\Xhline{1.5pt}
				First \Lame parameter & $\lambda$ & $118875.0$ & \si{\mega\pascal}\\
				Second \Lame parameter & $\mu$ & $79250.0$ & \si{\mega\pascal}\\
				Initial yield stress & $\tau_y$ & $308.26$ & \si{\mega\pascal}\\
				Kinematic hardening modulus & $H_a$ & $3774.25$ & \si{\mega\pascal}\\
				Kinematic yield stress saturation & $B_a$ & $175.55$ & -\\
				Saturation hardening stress & $\Delta \tau$ & $1.1761 \cdot 10^{6}$ & \si{\mega\pascal}\\
				Isotropic hardening parameter & $\kappa_u$ & $301.41$ & -\\
				\hline
				\multicolumn{4}{| c |}{Anisotropic (\modelIIa model)}\\
				\hline
				Damage modulus & $C$ & $1256.7$ & \si{\mega\pascal^{-1}}\\
				Damage exponent & $m$ & $0.2$ & -\\
				\hline
				\multicolumn{4}{| c |}{Isotropic (\modelIIb model)}\\
				\hline
				Damage modulus & $\hat{C}$ & $623.1$ & \si{\mega\pascal^{-1}}\\
				Damage exponent & $\hat{m}$ & $0.2$ & -\\
				\hline
			\end{tabular}
			\caption{Model parameters of the \modelII model for case-hardenend steel 16MnCrS5. Parameters associated with the isotropic damage model are indicated with a superposed hat $\hat\bullet$.}\label{Tab:Fundamentals:ModelParametersLemAniso}%
		\end{table}%

		All models match the experimental data well, see Fig.~\ref{Fig:Fits} (a, b). The largest error between the experiments and the calibration is associated with the \modelIIa model and occurs at a strain of $\ve_{11} = 0.07$. However, the respective error is below \SI{10}{\mega\pascal} and hence, less than $2.5 \%$. The behavior with respect to a shear mode is additionally shown in Fig.~\ref{Fig:Fits} (c) in order to highlight the models' different response -- the shear mode was not part of the calibration. All models show qualitatively similar behavior with a maximum deviation of less than $8 \%$. It shall be underlined that different models are used on purpose for a better and model-independent interpretation of the results. An evaluation of the individual models themselves is not intended and clearly depends on the intended application.
		
	\begin{remark}
		The constitutive models are evaluated at the material point level and calibrated to experiments with spatially constant stress and strain fields (no necking). Therefore, no regularization is required. If boundary value problems with localization are to be analyzed, a regularization is required in order to render finite element simulations well-posed. The respective energy contributions are, however, disregarded in this work, cf.~\cite{langenfeld_micromorphic_2020}.
	\end{remark}

    \subsection{Finite element simulations of a tension-torsion test}
    	
    	\begin{figure}[ht]
    		\centering
    		\psfrag{H}[l][l]{5}
    		\psfrag{W}[b][b]{2.5}
    		\psfrag{R}[b][b]{R200}
    		\includegraphics[width=0.5\textwidth]{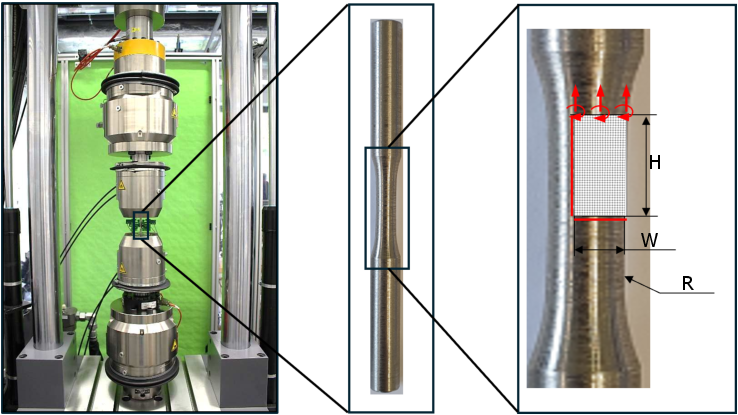}
    		\caption{Finite element simulation of a tension-torsion test: Servohydraulic Axial - Torsional Test System and the tensile specimen (case-hardened steel 16MnCrS5) with an overly of the finite element discretization of the simulated domain with 800 axisymmetric elements. \label{Fig:TensileSpecimenAndMachine}}
    	\end{figure}
    	
    	The goal of this subsection is to highlight the load path dependencies of material models for ductile damage. Therefore, the tensile specimen in Fig.~\ref{Fig:TensileSpecimenAndMachine} is loaded by means of two different load paths. The final deformation boundary conditions are identical in both cases, being \SI{0.2}{\milli \metre} in tensile direction and \SI{15}{\degree} in torsional direction. In one instance, the tensile load was applied prior to the torsional load, while in another instance, the order was reversed. As far as the simulation is concerned, only the rejuvenated part of the specimen is calculated and discretized by means of 800 elements. The kinematics are calculated with the assumption of axisymmetry and by means of logarithmic strains in line with the Green-Naghdi-theory, cf~\cite{GreenNaghdi1965}. The results of all four models (isotropic and anisotropic Ekh- and Lemaitre-model) are shown in Fig.~\ref{Fig:FEM_sim} and are given in terms of equivalent plastic strains and relative directional elastic stiffness at the final stage of deformation. Here, the relative directional elastic stiffness has been chosen as a first damage measure. Further damage measures are derived in detail in Subsec.~\ref{Ssec:Optimization}.
    	
		At first glance, the results seem to be path-independent. However, a more detailed analysis, i.e. the analysis of a single element, reveals a quantitative difference. The lower right element has been chosen as the element of interest for further investigations. Three particularities can accordingly be highlighted in the present finite-size analysis. First, the resulting damage differs by up to \SI{4.6}{\percent} between both load paths (\modelII model). The performance in terms of lifetime, for instance, can be very sensitive even to such differences in initial forming-induced damage (see~\cite{Langenfeld2023}). Second, equivalent plastic strain and damage do not always correlate positively (e.g., for the \modelI model). These findings highlight the need for analyzing and more importantly quantifying the load path dependence on damage evolution.
    		
    	In contrast to the current study, the numerical experiments of the following sections focus on the constitutive relations and are thus based on the analyses of single material points (local analyses).
   
   \afterpage{%
   	%
   	%
   	%
   	\begin{figure}[p!] %
   		\centering
   		\psfrag{A}[c][c]{\makebox{\modelI model}}
   		\psfrag{B}[c][c]{\makebox{\modelII model}}
   		\psfrag{C}[c][c]{\makebox{Anisotropic}}
   		\psfrag{D}[c][c]{\makebox{Isotropic}}
   		\psfrag{E}[c][c]{\makebox{Anisotropic}}
   		\psfrag{F}[c][c]{\makebox{Isotropic}}
   		\psfrag{G}[c][c]{\makebox{Equivalent plastic strain}}
   		\psfrag{H}[c][c]{\makebox{Relative directional elastic stiffness}}
   		\psfrag{J}[c][c]{\makebox{Tension $\rightarrow$ Torsion}}
   		\psfrag{K}[c][c]{\makebox{Torsion $\rightarrow$ Tension}}
   		\psfrag{M}[c][c]{\makebox{Tension $\rightarrow$ Torsion}}
   		\psfrag{N}[c][c]{\makebox{Torsion $\rightarrow$ Tension}}
   		\psfrag{S1} [c][l]{\makebox{\scalebox{0.6}{$0.138$}}} 
   		\psfrag{S2} [c][l]{\makebox{\scalebox{0.6}{$0.139$}}} 
   		\psfrag{S3} [c][l]{\makebox{\scalebox{0.6}{$0.152$}}} 
   		\psfrag{S4} [c][l]{\makebox{\scalebox{0.6}{$0.159$}}} 
   		\psfrag{S5} [c][l]{\makebox{\scalebox{0.6}{$0.143$}}} 
   		\psfrag{S6} [c][l]{\makebox{\scalebox{0.6}{$0.146$}}} 
   		\psfrag{S7} [c][l]{\makebox{\scalebox{0.6}{$0.139$}}} 
   		\psfrag{S8} [c][l]{\makebox{\scalebox{0.6}{$0.141$}}} 
   		\psfrag{S9} [c][l]{\makebox{\scalebox{0.6}{$0.747$}}} 
   		\psfrag{S10}[c][l]{\makebox{\scalebox{0.6}{$0.743$}}} 
   		\psfrag{S11}[c][l]{\makebox{\scalebox{0.6}{$0.753$}}} 
   		\psfrag{S12}[c][l]{\makebox{\scalebox{0.6}{$0.665$}}} 
   		\psfrag{S13}[c][l]{\makebox{\scalebox{0.6}{$0.752$}}} 
   		\psfrag{S14}[c][l]{\makebox{\scalebox{0.6}{$0.747$}}} 
   		\psfrag{S15}[c][l]{\makebox{\scalebox{0.6}{$0.772$}}} 
   		\psfrag{S16}[c][l]{\makebox{\scalebox{0.6}{$0.707$}}} 
   		\psfrag{U1}[l][l]{\scalebox{0.8}{\rotatebox{90}{\makebox{$1.6 \cdot 10^{-1}$}}}}
   		\psfrag{L1}[l][l]{\raisebox{2pt}{\rotatebox{90}{\scalebox{0.8}{\makebox{$4.2 \cdot 10^{-2}$}}}}}
   		\psfrag{T1}[c][c]{\scalebox{0.8}{\makebox{$\ve^{\text{p,eq}}$}}}
   		\psfrag{U2}[l][l]{\scalebox{0.8}{\rotatebox{90}{\makebox{$0.93$}}}}
   		\psfrag{L2}[l][l]{\raisebox{2pt}{\rotatebox{90}{\scalebox{0.8}{\makebox{$0.66$}}}}}
   		\psfrag{T2}[c][c]{\scalebox{0.8}{\makebox{$\xi_{\mathbb{E}}$}}}
   		\includegraphics[width=\textwidth]{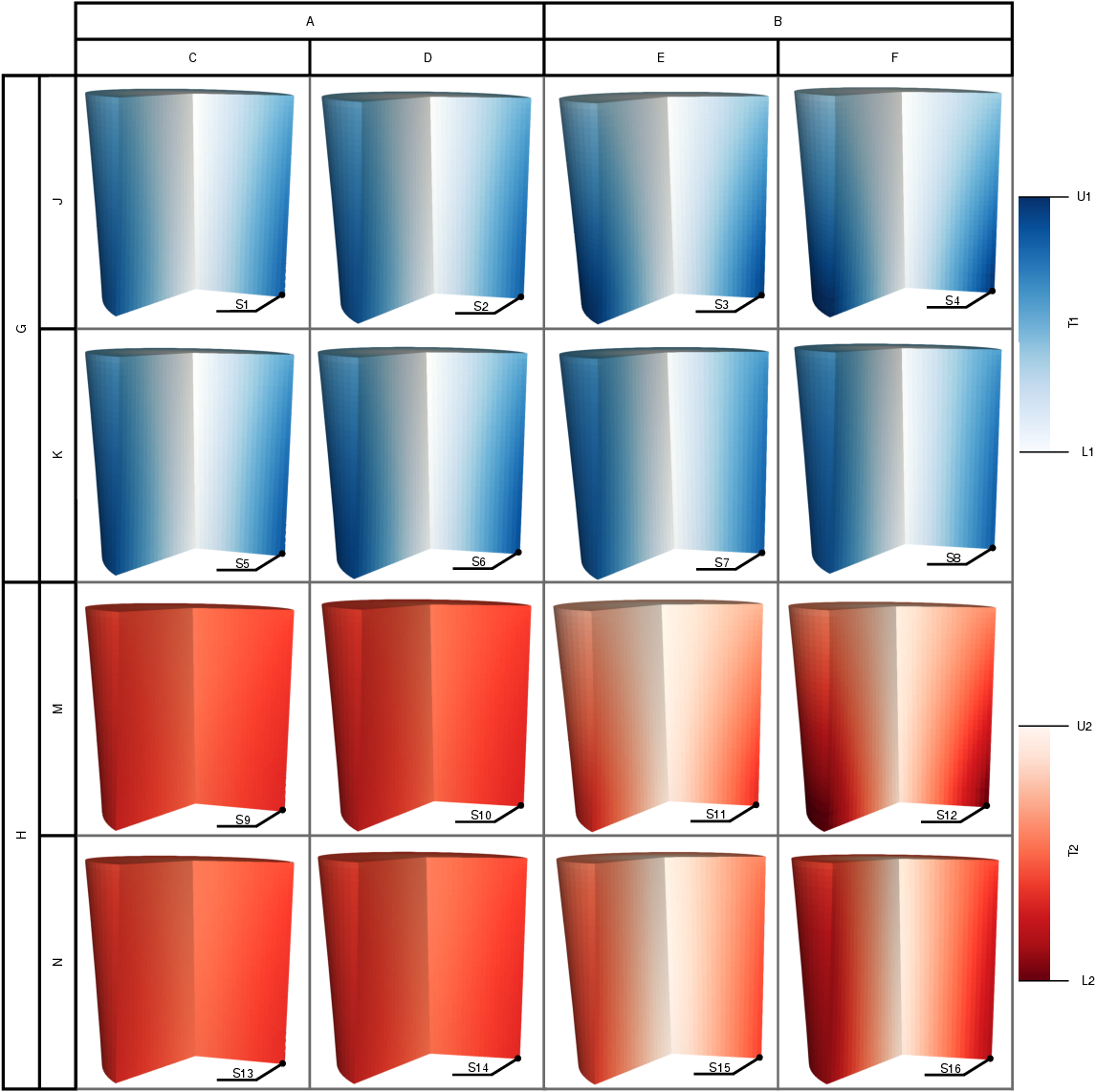}
   		\caption{Finite element simulation of a tension-torsion test: Equivalent plastic strain and relative directional elastic stiffness for two different load paths: first tension, then torsion; and first torsion, then tension. Results are shown for models \modelIa, \modelIb, \modelIIa and \modelIIb.}
   		\label{Fig:FEM_sim}
   	\end{figure}%
   }

\section{Methodology of evaluation: load path parametrization and damage measures}\label{Sec:Methodology}
	The analysis of varying load paths is a key of the present evaluation of damage descriptions. For example, the predictions of the previously introduced models will be compared along load paths of identical stress triaxialities and/or Lode angle parameters. For that reason, the parametrization of the load paths is explained in the following Subsection~\ref{Ssec:LPdef}. Characteristic damage measures are furthermore defined in Subsection~\ref{Ssec:Optimization} and based on elastic stiffness and elastic compliance tensors. Each of these measures allows to evaluate specific types of material degradation. For example, the Frobenius norm of the degraded elasticity tensor can be interpreted as an average stiffness measure. These characteristic measures will take the role of target functions for the load path analysis. They allow the comparison of load paths with respect to their damage impact, the identification of sensitivities and the exploration of missing influences.
	
	\subsection{Parametrization of load paths}\label{Ssec:LPdef}
		The numerical studies of the next section analyze the behavior of a representative material point. The mechanical loading of these points will hence be defined by prescribing either components of the stress or the strain tensor. Since these tensors are energetically dual, symmetric second-order tensors, one has to prescribe precisely six components in total. The dual six components follow as reactions. For instance, a strain-driven uniaxial tensile test corresponds to the following strain and stress tensors -- boxed components are prescribed, the other ones follow as reactions:
		\begin{align}\label{eq:stress-strain-components}
			\B{\ve}= \begin{bmatrix}
				\boxed{\ve_{11}} && \ve_{12} && \ve_{13}
				\\
				\text{sym}  && \ve_{22} &&  \ve_{23}
				\\
				\text{sym} && \text{sym} && \ve_{33}
			\end{bmatrix}_{\left[\B{e}_1, \B{e}_2, \B{e}_{3} \right]}
			\qquad
			\B{\sigma}= \begin{bmatrix}
				\sigma_{11} && \boxed{\sigma_{12}} && \boxed{\sigma_{13}}
				\\
				\text{sym}  && \boxed{\sigma_{22}} &&  \boxed{\sigma_{23}}
				\\
				\text{sym} && \text{sym} && \boxed{\sigma_{33}}
			\end{bmatrix}_{\left[\B{e}_1, \B{e}_2, \B{e}_{3} \right]}
		\end{align}
		In Eq.~\eqref{eq:stress-strain-components}, the components are defined with respect to a chosen basis $\B{e}_i$ (here, a Cartesian setting is adopted). Loading is implemented by enforcing a strain in $11$-direction for this example of a uniaxial tensile test. Hence, the respective components of the stress tensor are unknown, but follow from the constitutive model. The other components of the stress tensor have to vanish, prescribing $\sigma_{ij}=0$ except for $\sigma_{11}$.
		
		The evolution of prescribed components is either parametrized and discretized by piecewise linear functions of the form 
		\begin{align}
			f_\mr{pwl}\of{t; \B{P}} &= P_i + \frac{P_i - P_j}{t_i - t_j} \, t \quad \text{if } t \in \left[t_i, \, t_j\right] \, \text{,}
		\end{align}
		or by \Bezier curves defined as
		\begin{align}\label{eq:bezier}
			f_\mr{Bez}\of{t; \B{Q}} &= \sum_{i=0}^{N} {N \choose i} \, \left[1 - t\right]^{\left[N-i\right]} \, t^i \, Q_i \, \text{,}
		\end{align}
		with control points \( P_i \) and \( Q_i \), respectively. The time is normalized within the interval $t\in[0;1]$, where the time-scaling is irrelevant since rate-independent models are considered.
	
		From the perspective of the numerical implementation, prescribing components of the stress tensor requires an iteration. This is implemented by a constitutive driver where the respective constraints are solved by means of a Newton-iteration. An analogous procedure is also employed for controlling the stress triaxiality and/or the Lode (angle) parameter.
	
	\subsection{Characteristic damage measures and target functions}\label{Ssec:Optimization}
		Characteristic damage measurements are defined in order to compare the damage evolution along the load paths and between different models. Since the internal variables capturing damage are not identical for both basic models, the respective elastic stiffness and elastic compliance tensors are instead considered for the definition of suitable damage measures. Elastic stiffness tensor and elastic compliance tensors are inverse to each other, i.e., $\mathbb{E}\el:\mathbb{C}\el=\mathbb{I}$ with $\mathbb{I}$ being the symmetric fourth-order identity. Starting from the Helmholtz energy $\psi$, the elastic moduli of the \modelI model are obtained as
		\begin{align}
			\left[\mathbb{E}_\mr{\modelI}\el\right]_{ijkl}:=\frac{\partial^2\psi}{\partial\ve_{ij}\partial\ve_{\KF{kl}}}= \lambda \, \left[b_{ij} \, b_{kl}\right] + \mu \left[b_{ik} \, b_{jl} + b_{il} \, b_{jk}\right] \, \text{,}
		\end{align}
		with $\mathbb{E}_\mr{\modelI}\el:\mathbb{C}_\mr{\modelI}\el=\mathbb{I}$. Likewise, the \modelII model yields the elastic compliance
		\begin{align}
			\begin{split}
				\left[\mathbb{C}_\mr{\modelII}\el\right]_{ijkl}:=\frac{\partial^2\mathcal{G}}{\partial\sigma_{ij}\partial\sigma_{kl}}
				=& \frac{1+\nu}{E} \, \left[\frac{1}{2} \, \left[ H_{ik} \, H_{jl} + H_{il} \, H_{jk}\right] - \frac{1}{3} \, \left[ \delta_{ij} \, H_{km} \, H_{ml} + H_{in} \, H_{nj} \, \delta_{kl} \right]\right.
				\\
				&\phantom{\coloneqq \frac{1+\nu}{E} \,} + \left. \dfrac{1}{9} \, \left[\, H_{op} \, H_{op}  + \frac{3}{1 - D^\mr{hyd}} \right] \, \delta_{ij} \, \delta_{kl}\right] - \frac{\nu}{E} \, \frac{1}{1 -  D^\mr{hyd}} \, \delta_{ij} \, \delta_{kl} \, \text{,}
			\end{split} 
		\end{align}
		where $\delta_{ij}$ is the Kronecker-Delta and $\mathbb{E}_\mr{\modelII}\el:\mathbb{C}_\mr{\modelII}\el=\mathbb{I}$. 
		
		Effective-scalar-valued variables can be introduced based on either the elastic stiffness or the elastic compliance tensor. For instance, the Frobenius-norm of the fourth-order elasticity tensor
		\begin{align}
			f_\mathbb{E} \coloneqq \sqrt{\mathbb{E}\el :: \mathbb{E}\el}\, \text{,} \label{Eq:target}
		\end{align}
		with $::$ denoting a quadruple contraction, defines a suitable average value of $\mathbb{E}\el$. Based on Eq.~\eqref{Eq:target} and denoting the elasticity tensor associated with a reference model or load path as $\bar{\mathbb{E}}$, the difference between two load paths (or model descriptions) can be defined as
		\begin{align}
			f_\mathbb{E}^\mathrm{diff} &= \sbr{\sqrt{\mathbb{E}\el :: \mathbb{E}\el} - \sqrt{\bar{\mathbb{E}}\el :: \bar{\mathbb{E}}\el}}^2 \label{Eq:target2}\,.
		\end{align}

		The undamaged configuration represents a natural reference solution, i.e., $\bar{\mathbb{E}}\el=\mathbb{E}\el(t=0)$.
		As an alternative to the Frobenius-norm~\eqref{Eq:target}, the directional elastic stiffness and compliance
		\begin{align}
			E_{\B{r}} = \left[\B{r} \otimes \B{r} \right] : \mathbb{E}\el : \left[\B{r} \otimes \B{r} \right]\, \text{,} \label{Eq:directional-E}
			\\
			C_{\B{r}} = \left[\B{r} \otimes \B{r} \right] : \mathbb{C}\el : \left[\B{r} \otimes \B{r} \right]\, \text{,}
			\label{Eq:directional-C}
		\end{align}
		can be considered respectively, where unit vector $\B{r}$ defines the direction. A suitable measure for the damage accumulation based on Eq.~\eqref{Eq:directional-E} reads, for instance,
		\begin{align}
			\xi_\mathbb{E} &= \frac{ \min_{\B{r}} E_{\B{r}}\of{\B{r}}}{E_0} \label{Eq:Opt:xi_d}\,.
		\end{align}
		Here, $E_0$ corresponds to the initial state at $t=0$, which is isotropic and independent of $\B{r}$. In contrast to measure~\eqref{Eq:target2}, measure~\eqref{Eq:Opt:xi_d} depends on the directional-dependent lowest stiffness and thus largest damage-induced degradation. Finally, if the direction-dependent elastic modulus is to be evaluated, \( C_{\boldsymbol{r}} \) can be compared to the reference resulting in
		\begin{align}
			\xi_{\mathbb{C}}(\boldsymbol{r}) = \dfrac{C_0}{C_{\boldsymbol{r}}}\,. \label{eq:st:stiffness_ratio}
		\end{align}


\section{Numerical examples}\label{Sec:NumericalResults}
	This section highlights the application limits of isotropic damage representations by means of illustrative examples. These examples are structured by the re-evaluation of well-established statements on damage characterization and control. This allows a connection to practically relevant load path domains and to better identify influences that have not yet been considered in full depth. In contrast to previous studies, the accuracy of the different damage representations is also analyzed quantitatively. First, the well-known statement
	\begin{itemize}
		\item The smaller the stress triaxiality, the smaller the damage accumulation\\(see~\cite{tekkaya_forming-induced_2017} and Subsection~\ref{Ssec:TensShear})
	\end{itemize}
	as well as
	\begin{itemize}
		\item Damage accumulation is uniquely governed by the stress triaxiality and the Lode (angle) parameter (see~\cite{andrade_incremental_2016,zhu_parameters_2023})
	\end{itemize}
	are analyzed and put into a quantitative context. It will be shown that contradictory load paths can indeed be designed by means of optimization algorithms. Second, the concept of fracture surfaces in line with~\cite{wierzbicki_new_2005,BAI20081071} is investigated in Subsection~\ref{Ssec:FractureSurfaces}. This concept is based on the statement
	\begin{itemize}
		\item Damage accumulation is uniquely governed by the triple stress triaxiality, Lode (angle) parameter and equivalent plastic strain (see~\cite{basaran_stress_2011,andrade_incremental_2016,neukamm_lokalisierung_2018} and Subsection~\ref{Ssec:FractureSurfaces}).
	\end{itemize}
	Optimization algorithms again show that this statement does also not hold true in general -- particularly for complex load paths. Since the present studies rely on numerical constitutive modeling, the two different prototype models \textit{\modelI} and \textit{\modelII} with isotropic and anisotropic variants are considered throughout the discussion.
	
	\subsection{The smaller the stress triaxiality, the smaller the damage accumulation? \\Coupled tension-torsion experiments}\label{Ssec:TensShear}
\begin{figure}[ht]
			\centering
			\hfill
			\begin{subfigure}[ht]{0.3\textwidth}
				\centering
				\psfrag{uz}[c][c]{$u$}
				\psfrag{gz}[c][c]{$\gamma$}
				\psfrag{g}[c][c]{$\theta$}
				\psfrag{r}[c][c]{$R$}
				\psfrag{l}[c][c]{$L$}
				\psfrag{e1}[c][c]{$\B{e}_r$}
				\psfrag{e2}[c][c]{$\B{e}_\theta$}
				\psfrag{e3}[c][c]{$\B{e}_z$}
				\psfrag{e12}[c][c]{$\theta,\!z$}
				\includegraphics[width=0.75\textwidth]{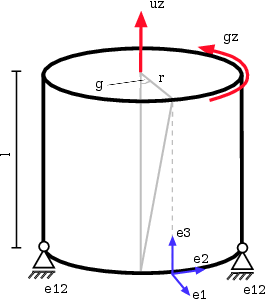}
				\hspace{60pt}
			\end{subfigure}
			\hfill
			\begin{subfigure}[ht]{0.3\textwidth}
				\centering
				\begin{align*}
					\B{\ve} &= \begin{bmatrix}
						\ve_{rr} && \ve_{r\theta} && \ve_{rz}
						\\
						\text{sym}  && \ve_{\theta \theta} &&  \boxed{\ve_{\theta z}}
						\\
						\text{sym} && \text{sym} && \boxed{\ve_{zz}}
					\end{bmatrix}_{\left[\B{e}_\mr{r}, \B{e}_\theta, \B{e}_{z} \right]}
					\\
					\B{\sigma} &= \begin{bmatrix}
						\boxed{\sigma_{rr}} && \boxed{\sigma_{r\theta}} && \boxed{\sigma_{rz}}
						\\
						\text{sym}  && \boxed{\sigma_{\theta\theta}} &&  \sigma_{\theta z}
						\\
						\text{sym} && \text{sym} && \sigma_{zz}
					\end{bmatrix}_{\left[\B{e}_\mr{r}, \B{e}_\theta, \B{e}_{z} \right]}
				\end{align*}
			\end{subfigure}
			\hspace{2cm}
			\caption{Combined axial/torsional test: sketch of mechanical system and boundary conditions in terms of prescribed stress and strains. Prescribed coordinates are boxed. According to the underlying kinematics, $\ve_{zz} = u/L$ and $\ve_{\theta z} = 1/2 \, \gamma \, R/L$.}
			\label{Fig:ST:Scheme}
		\end{figure}

		Inspired by experimental axial-torsional test systems and with the goal of multi-directional load paths in mind, a tension-torsion problem is considered within a cylindrical coordinate system $\left[\B{e}_\mr{r}, \B{e}_\theta, \B{e}_{z} \right]$, cf.~Fig.~\ref{Fig:ST:Scheme}.
		A virtual tension-torsion sequence serves as the reference load path. Within this test, torsional strain $\ve_{\theta z}$ is increased first up to $\ve_{\theta z} = 0.03$ (while $\ve_{zz} = 0= \text{const.}$), and subsequently tensile strain $\ve_{zz}$ is increased up to $\ve_{zz} = 0.06$ (while $\ve_{\theta z} = 0.03 = \text{const.}$). The final strain components at normalized time $t=1$ are  $\ve_{zz} = 0.06$ and $\ve_{\theta z} = 0.03$. The evolution of the stress-triaxiality corresponding to this reference load path is denoted by $\eta^{\text{ref}}(t)$ and the Frobenius-norm of the final elastic stiffness tensor is $f_{\mathbb{E}}^{\text{ref}}(t=1)$. Since four different models are analyzed, four different reference solutions are computed (isotropic and anisotropic \modelI model, isotropic and anisotropic \modelII model).
		
		Next, load path variations are parametrized to search for damage-mitigating alternatives with higher stress triaxialities. These would then provide counter examples for the first statement ("The smaller the stress triaxiality, the smaller the damage accumulation"). Strain components $\ve_{\theta z}$ and $\ve_{zz}$ are \KF{each} discretized by means of fifth-order \Bezier curves, see Eq.~\eqref{eq:bezier}. While the initial conditions are $\ve_{\theta z}(t=0)=\ve_{zz}(t=0)=0$, the final state is defined by $\ve_{\theta z}(t=1)=0.03$ and $\ve_{zz}(t=1)=0.06$. The \KF{remaining eight} coordinates of the \Bezier representation are computed by mitigating the final damage state in the sense of an optimization. To be more precise, the following optimization problem is considered:
		\begin{align}
			\underset{Q_i}{\text{max}} \, f_\mathbb{E}\of{t=1} \quad \text{subject to} \,\,  \eta^\mr{opt}(t) \geq \eta^\mr{ref}(t) \, \forall \, t \in \left[0,  1\right] \, \text{.}
		\end{align}
		Accordingly, we search for the load path in terms of control points \( Q_i\) that maximizes the average elastic stiffness. This path has to fulfill the constraint $\eta^\mr{opt}(t) \geq \eta^\mr{ref}(t) \, \forall \, t$. Since the resulting optimization problem is highly non-convex, a two-stage gradient-free optimization method is employed. To be more explicit, a particle swarm optimization (see~\cite{488968}), followed up by a Nelder-Mead simplex algorithm (see~\cite{Nelder1965ASM}) is employed. The inequality $\eta^\mr{opt}(t) \geq \eta^\mr{ref}(t) \, \forall \, t$ is enforced by means of a penalty function.
		
\begin{figure}[ht]%
			\centering
			\begin{subfigure}[ht]{0.25\textwidth}
				\centering
				\psfrag{xi}[c][c]{\scalebox{0.7}{Relative stiffness $\xi_\mathbb{E} \, [-]$}}
				\psfrag{1Ia}[c][c]{\scalebox{0.6}{\modelIa}}
				\psfrag{2Ib}[c][c]{\scalebox{0.6}{\modelIb}}
				\psfrag{3IIa}[c][c]{\scalebox{0.6}{\modelIIa}}
				\psfrag{4IIb}[c][c]{\scalebox{0.6}{\modelIIb}}
				\begin{overpic}[width=0.98\textwidth]{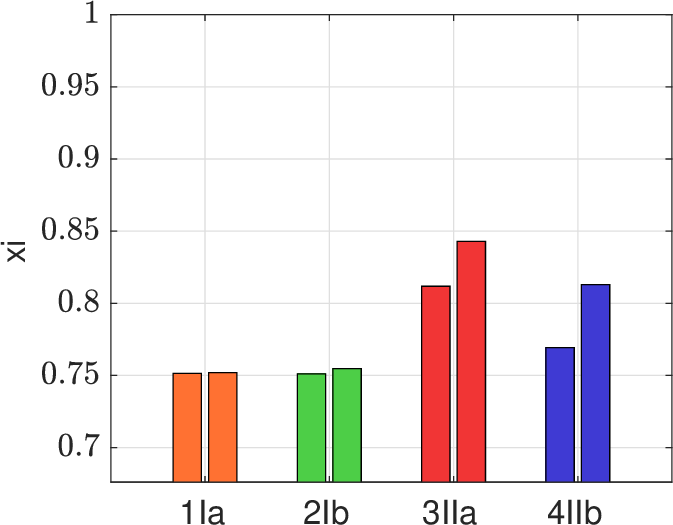}
					\put(25, 23){\rotatebox{90}{\scriptsize Reference}}
					\put(31, 23){\rotatebox{90}{\scriptsize Optimized}}
				\end{overpic}
				\setlength{\abovecaptionskip}{5.5pt}
				\caption{Final relative stiffness $\xi_\mathbb{E}$.~\\\phantom{.}}%
			\end{subfigure}
			\hspace{0.1cm}
			\begin{subfigure}[ht]{0.25\textwidth}
				\centering
				\vspace{0.07cm}
				\psfrag{p}[c][c]{\raisebox{0.2cm}{\scalebox{0.7}{Equiv. plastic strain \PEEQ $\left[-\right]$}}}
				\psfrag{1Ia}[c][c]{\scalebox{0.6}{\modelIa}}
				\psfrag{2Ib}[c][c]{\scalebox{0.6}{\modelIb}}
				\psfrag{3IIa}[c][c]{\scalebox{0.6}{\modelIIa}}
				\psfrag{4IIb}[c][c]{\scalebox{0.6}{\modelIIb}}
				\includegraphics[width=0.98\textwidth]{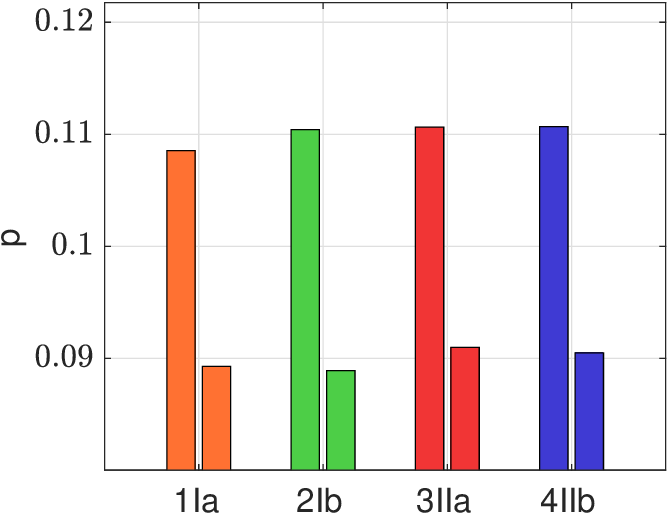}
				\setlength{\abovecaptionskip}{5.5pt}
				\caption{Final equivalent plastic strain \PEEQ.}%
			\end{subfigure}
			\hspace{0.1cm}
			\begin{subfigure}[ht]{0.25\textwidth}
				\psfrag{stiffness I-a}[c][c]{\scalebox{0.7}{Relative stiffness $\xi_\mathbb{E} \, \left[-\right]$}}
				\psfrag{p}[c][c]{{\scalebox{0.7}{Equivalent plastic strain \PEEQ $\left[-\right]$}}}
				\includegraphics[width=0.98\textwidth]{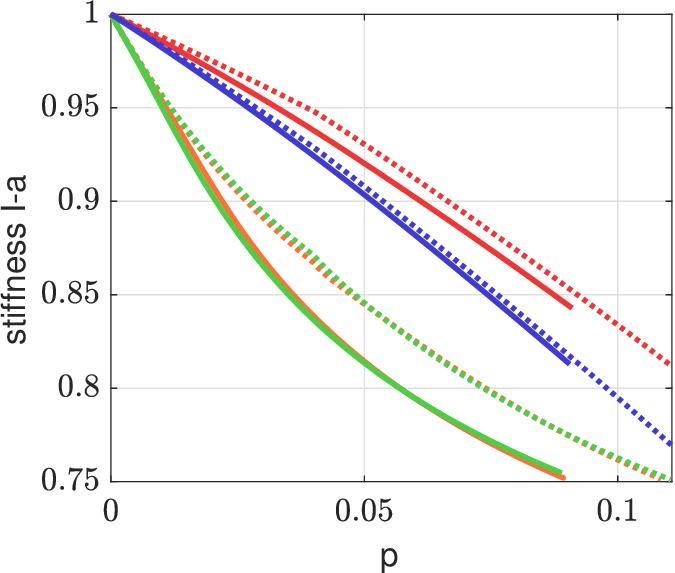}
				\setlength{\abovecaptionskip}{9pt}
				\caption{Relative stiffness $\xi_\mathbb{E}$ against equivalent plastic strain \PEEQ.}%
			\end{subfigure}
			\hspace{0cm}
			\begin{subfigure}[ht][80pt][t]{0.07\textwidth}
				\centering
				\psfrag{A}[l][l]{\scalebox{0.7}{\modelIa ref.}}
				\psfrag{B}[l][l]{\scalebox{0.7}{\modelIb ref.}}
				\psfrag{C}[l][l]{\scalebox{0.7}{\modelIIa ref.}}
				\psfrag{D}[l][l]{\scalebox{0.7}{\modelIIb ref.}}
				\psfrag{E}[l][l]{\scalebox{0.7}{\modelIa opt.}}
				\psfrag{F}[l][l]{\scalebox{0.7}{\modelIb opt.}}
				\psfrag{G}[l][l]{\scalebox{0.7}{\modelIIa opt.}}
				\psfrag{H}[l][l]{\scalebox{0.7}{\modelIIb opt.}}
				\vspace{-0.65cm}
				\includegraphics[width=0.85\textwidth]{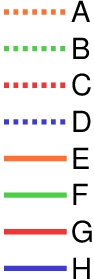}
			\end{subfigure}
			\hspace{1.6cm}
			\caption{Combined axial/torsional test: (a) Final relative stiffness $\xi_\mathbb{E}(t=1)$ and (b) final equivalent plastic deformation \PEEQ$(t=1)$ for all four constitutive models and for the reference load path (left bars) as well as for the optimized load path (right bars). (c) Evolution of effective relative stiffness $\xi_\mathbb{E}$ in terms of \PEEQ.}
			\label{Fig:ST:xi_bar}
		\end{figure}

		The results in Fig.~\ref{Fig:ST:xi_bar} correspond to eight different simulations -- four constitutive models (isotropic and anisotropic \modelI model, isotropic and anisotropic \modelII model) times two load paths (reference path and optimized path). 
		It can be seen that the optimized paths indeed lead to less damage, i.e. their average elastic stiffness is larger (right columns in Fig.~\ref{Fig:ST:xi_bar} (a)). Since ductile damage is primarily driven by plastic deformations, one might expect that \PEEQ is smaller for the optimized paths. This is confirmed for the present scenario in Fig.~\ref{Fig:ST:xi_bar}: higher relative stiffness in Fig. Fig.~\ref{Fig:ST:xi_bar} (a) corresponds to less equivalent plastic strain in Fig.~\ref{Fig:ST:xi_bar} (b). For this reason, a low equivalent plastic strain is suggested as a further damage-mitigating indicator. This hypothesis can also be drawn from the monotonically decreasing courses in Fig.~\ref{Fig:ST:xi_bar} (c). To be more specific, Figs.~\ref{Fig:ST:xi_bar} (a, b) are end-point projections of the curves presented in Fig.~\ref{Fig:ST:xi_bar} (c) onto the $\xi_{\mathbb{E}}$-axis and the \PEEQ-axis, respectively. Although the curves associated with optimal paths are below their respective reference (i.e. less remaining stiffness at distinct values of \PEEQ), the final remaining stiffness can be increased by reducing the path length in \PEEQ. Therefore, the equivalent plastic strain \PEEQ is an additional important measure to characterize ductile damage, alongside of the stress triaxiality and the Lode angle parameter. Its influence will be examined in the next subsection.
		
\begin{figure}[ht]%
			\centering
			\begin{subfigure}[ht]{0.27\textwidth}
				\centering
				\psfrag{Strain11}[c][c]{\scalebox{0.8}{Tensile strain $\ve_{zz}$}}
				\psfrag{Strain12}[c][c]{\scalebox{0.8}{Shear strain $\ve_{\theta z}$}}
				\vspace{0.07cm}
				\begin{overpic}[width=\textwidth]{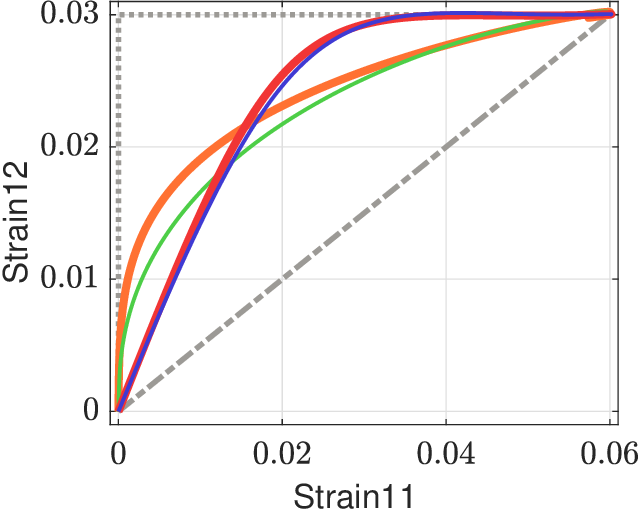}
					\put(20, 72.5){\scriptsize Reference}
					\put(45, 54){\rotatebox{28}{\scriptsize Optimized}}
					\put(45, 32){\rotatebox{39}{\scriptsize Start}}
				\end{overpic}
				\caption{Strain paths}
			\end{subfigure}
			\hspace{0.1cm}
			\begin{subfigure}[ht]{0.25\textwidth}
				\centering
				\psfrag{Time}[c][c]{\scalebox{0.8}{Normalized time $t \, \left[-\right]$}}
				\psfrag{triax opt}[c][c]{\scalebox{0.8}{Stress triaxiality $\eta \, \left[-\right]$}}
				\includegraphics[width=\textwidth]{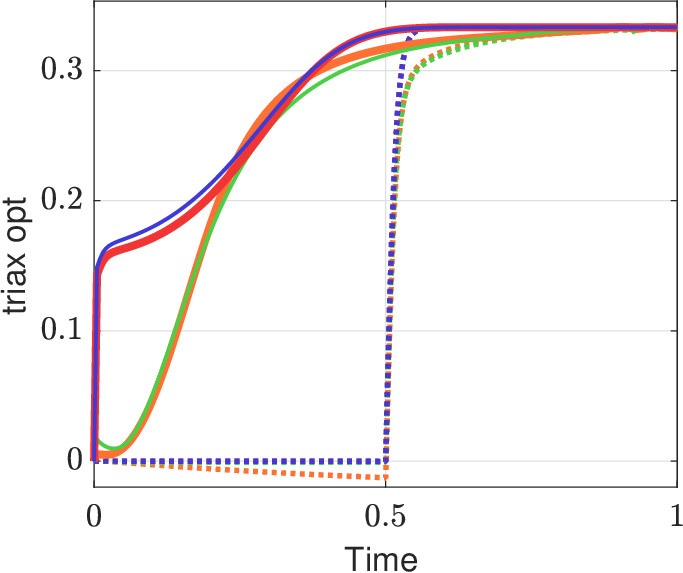}
				\caption{Stress triaxiality}
			\end{subfigure}
			\hspace{0.1cm}
			\begin{subfigure}[ht]{0.25\textwidth}
				\centering
				\psfrag{Time}[c][c]{\scalebox{0.8}{Normalized time $t \, \left[-\right]$}}
				\psfrag{p}[t][c]{\scalebox{0.8}{Cumul. eq. pl. strain $p \, \left[-\right]$}}
				\psfrag{lode I-b}[c][c]{\scalebox{0.8}{Lode angle par. $\bar{\theta} \, \left[-\right]$}}
				\includegraphics[width=\textwidth]{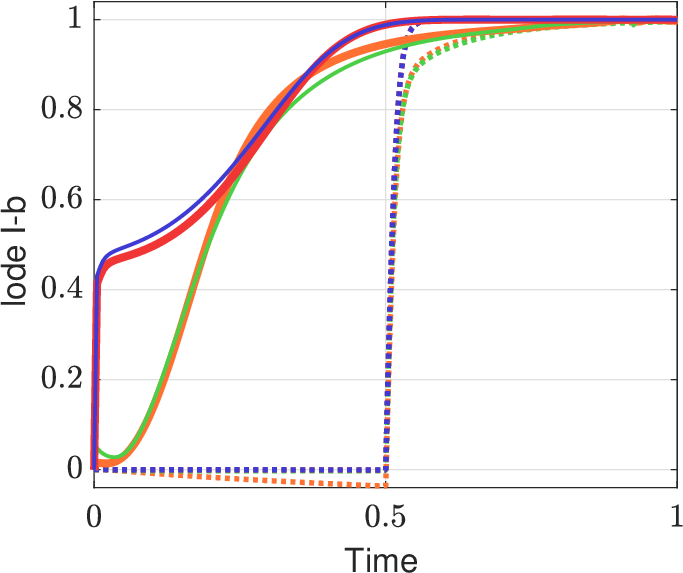}
				\caption{Lode angle parameter}
			\end{subfigure}
			\hspace{0cm}
			\begin{subfigure}[ht][80pt][t]{0.07\textwidth}
				\centering
				\psfrag{A}[l][l]{\scalebox{0.7}{\modelIa ref.}}
				\psfrag{B}[l][l]{\scalebox{0.7}{\modelIb ref.}}
				\psfrag{C}[l][l]{\scalebox{0.7}{\modelIIa ref.}}
				\psfrag{D}[l][l]{\scalebox{0.7}{\modelIIb ref.}}
				\psfrag{E}[l][l]{\scalebox{0.7}{\modelIa opt.}}
				\psfrag{F}[l][l]{\scalebox{0.7}{\modelIb opt.}}
				\psfrag{G}[l][l]{\scalebox{0.7}{\modelIIa opt.}}
				\psfrag{H}[l][l]{\scalebox{0.7}{\modelIIb opt.}}
				\vspace{-0.65cm}
				\includegraphics[width=0.9\textwidth]{Legend2.eps}
			\end{subfigure}
			\hspace{1.6cm}
			\caption{Combined axial/torsional test: Reference and optimized load paths for different ductile damage models. Reference paths are represented by dotted lines, optimized load paths by solid lines.}
			\label{Fig:ST:paths}
		\end{figure}

		According to Fig.~\ref{Fig:ST:paths} (b), the stress triaxiality of the optimized paths is always higher than that of the reference paths, i.e., inequality $	\eta^\mr{opt}(t) \geq \eta^\mr{ref}(t)$ is indeed fulfilled at any time. 
		As evident from Fig.~\ref{Fig:ST:paths} (a), shear component $\ve_{\theta z}$ and axial component $\ve_{zz}$ of the strain tensor are simultaneously active for the optimal paths -- in contrast to the sequential reference path. Finally, the evolution of the Lode angle parameter is shown in Fig.~\ref{Fig:ST:paths} (c). \KF{Due to plane stress conditions, the stress tensor shows only two non-vanishing eigenvalues. In this case, the Lode angle parameter can be expressed in terms of the stress triaxiality ($\bar{\theta} = 1 - \tfrac{2}{\pi}\,\arccos\of{\tfrac{3}{2}\,\eta\,\left[1-\tfrac{1}{3}\eta^2\right]}$) and, thus, both invariants show a similar behavior.} The present experiment includes no means to control the Lode angle parameter. However, literature indicates that it is an important indicator for ductile damage, along with stress triaxiality. Therefore, the Lode angle parameter will also be controlled in subsequent experiments.
		
\begin{figure}[h]
  \begin{minipage}[b]{0.28\linewidth}
    \psfrag{e1}[c][c]{$\B{e}_\mr{z}$}
    \psfrag{e2}[c][c]{$\B{e}_\theta$}
    \psfrag{e3}[c][c]{$\B{e}_\mr{r}$}
    \includegraphics[width=0.6\linewidth]{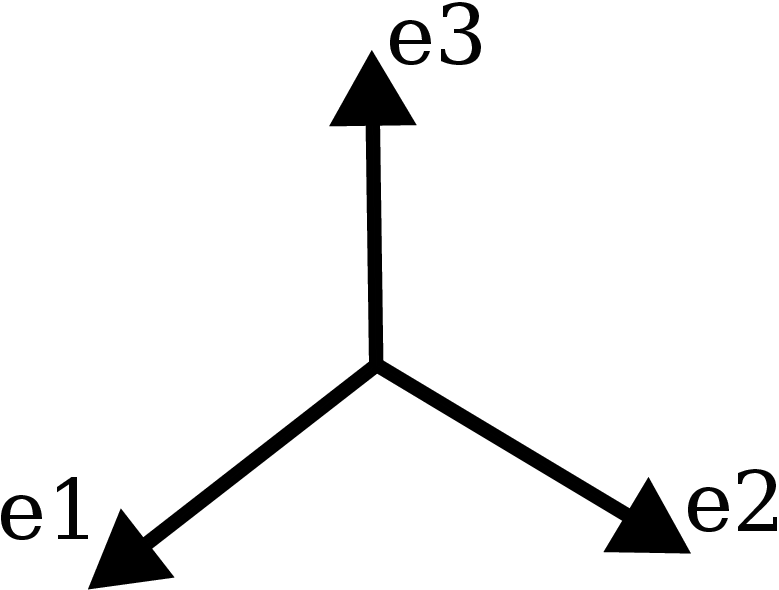}
    \vspace{2cm}
  \end{minipage}
  \hspace{-1cm}
  \begin{minipage}[b]{0.28\linewidth}
    \psfrag{A}[c][c]{\textcolor{white}{Ref.}}
	\psfrag{B}[c][c]{\textcolor{white}{Opt.}}
	\includegraphics[width=\textwidth]{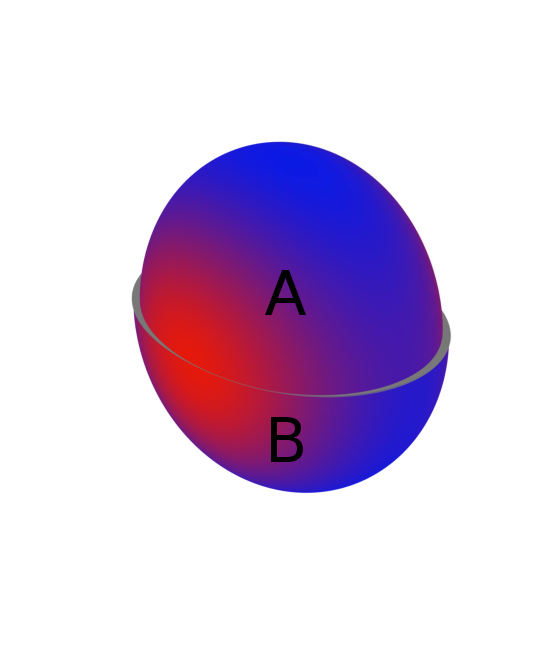}%
	\subcaption{Anisotropic damage model \modelIa}
  \end{minipage}
  \hfill
  \begin{minipage}[b]{0.28\linewidth}
    \psfrag{A}[c][c]{\textcolor{white}{Ref.}}
	\psfrag{B}[c][c]{\textcolor{white}{Opt.}}
	\includegraphics[width=\textwidth]{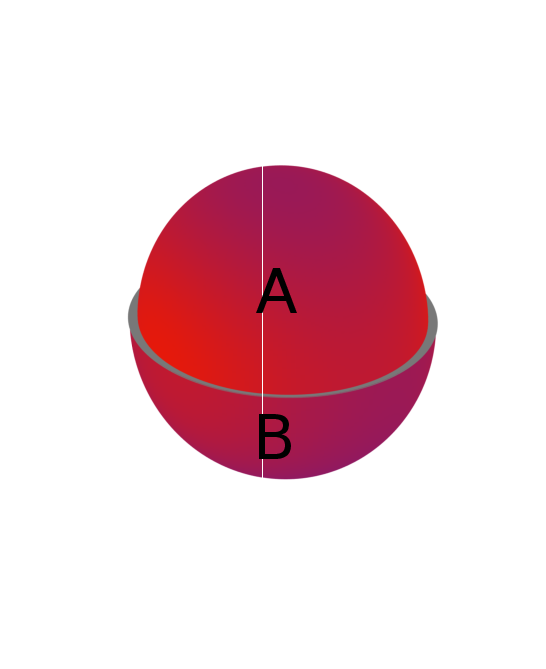}%
	\subcaption{Anisotropic damage model \modelIIa}
  \end{minipage}
  \hfill
  \begin{minipage}[b]{0.1\linewidth}
    \psfrag{E1}[c][c]{\makebox{\raisebox{4pt}{$\xi_{\mathbb{C}}$}}}
    \includegraphics[width=0.7\linewidth]{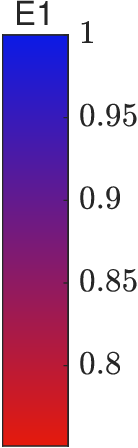}
    \vspace{1.5cm}
  \end{minipage}
\caption{Combined axial/torsional test: Directional relative stiffness \KF{according to Equation \eqref{eq:st:stiffness_ratio}} for anisotropic damage evolution. \KF{The vector $\B{r}$ in Equation~\eqref{Eq:directional-C} has been parametrized using spherical coordinates.} Comparison between optimized (lower) and reference paths (upper hemispheres) for $t=1$.}\label{Fig:ST:stiffness}
\end{figure}

		A more detailed comparison of the material degradation at the final time $t=1$ is given in Fig.~\ref{Fig:ST:stiffness}. It shows the directional relative stiffness of the anisotropic damage models \modelIa and \modelIIa relative to the initial isotropic, undamaged state. While the model \modelIIa leads to an almost isotropic material degradation, the model \modelIa predicts a more pronounced anisotropy. It shall be noted again that the comparison of the models does not aim at their individual evaluation but rather at finding model-invariant conclusions on the damage evolution and the corresponding input parameters. For both constitutive models, the optimized load paths are characterized by a larger elastic stiffness compared to the reference solution. Stress triaxiality -- even in combination with the equivalent plastic strain -- hence remains an insufficient indicator for damage.
		
From a methodological perspective, the computational framework allows to quantify the limits of the established isotropic damage description based on load path optimization.

	\subsection{Is damage accumulation uniquely governed by the triple stress triaxiality, Lode (angle) parameter and equivalent plastic strain?}\label{Ssec:FractureSurfaces}
		It has been shown in the previous section that the stress triaxiality as well as the equivalent plastic strain are primary influencing factors of ductile damage evolution. However, it has also been shown that these factors are not sufficient in order to characterize anisotropic material degradation under complex load paths. For this reason, the Lode angle parameter is often also considered. So-called {\em fracture surfaces} are a frequently employed damage framework that is based on the stress triaxiality, the Lode angle parameter and the equivalent plastic strain, cf.~\cite{wierzbicki_new_2005,BAI20081071}. More precisely, the damage evolution is then assumed to be proportional to the plastic strain rate and a scaling factor. The latter is a function in terms of the triple stress triaxiality, Lode angle parameter and the equivalent plastic strain. 
		Fracture surfaces, such as those in~\cite{wierzbicki_new_2005,BAI20081071}
		, can be interpreted as iso-surfaces of an equivalent scalar-valued damage measure. Accordingly, they can be computed for a given model by connecting triples of stress triaxiality, Lode angle parameter and the equivalent plastic strain to the same equivalent damage measure. First, such surfaces are computed in paragraph~\ref{subsec:proportional} for proportional load paths. Subsequently, the influence of more complex load paths is studied in paragraph~\ref{subsec:nonproportional} as well as in paragraph~\ref{subsec:nonproportional-2}. It will be shown in this section that even this ansatz is still not sufficient for a full characterization of ductile damage accumulation.
		
		\subsubsection{Fracture surfaces for proportional loading}\label{subsec:proportional}
			In order to compute iso-surfaces of damage, the directional relative stiffness of $\xi_\mathbb{E} = 0.8$ is chosen. Starting from the initial unloaded configuration, the stresses are increased until threshold $\xi_\mathbb{E} = 0.8$ is reached. For the whole load path, the stress triaxiality and the Lode angle parameter are kept constant (proportional loading). $17\times 13$ tuples of stress triaxiality and Lode angle parameter are considered in order to span the fracture surface. Along the evolving load path, the stress tensor is prescribed as
			\begin{align}\label{eq:stress-strain-components-proportional}
				\B{\sigma}= \begin{bmatrix}
					\boxed{\sigma_{11}} && \boxed{0} && \boxed{0}
					\\
					\text{sym}  && \boxed{\sigma_{22}(\sigma_{11},\eta,\bar{\theta})} &&  \boxed{0}
					\\
					\text{sym} && \text{sym} && \boxed{\sigma_{33}(\sigma_{11},\eta,\bar{\theta})}
				\end{bmatrix}_{\left[\B{e}_1, \B{e}_2, \B{e}_{3} \right]} \, \text{,}
			\end{align}
			where $\eta$ is the (prescribed) stress triaxiality and $\bar{\theta}$ the (prescribed) Lode angle parameter. Closed-form solutions for $\sigma_{22}(\sigma_{11},\eta,L\of{\bar{\theta}})$ and $\sigma_{33}(\sigma_{11},\eta,L\of{\bar{\theta}})$ are given in~\ref{Ssec:PR}. Thus, the only free loading parameter is $\sigma_{11}$. This parameter is increased until the damage threshold is reached.
			
\begin{figure}[ht]%
				\centering
				\hfill
				\psfrag{E}[c][c]{\scalebox{0.8}{Equivalent plastic strain \PEEQ [-]}}
				\begin{subfigure}[ht]{\textwidth}
					\begin{subfigure}[ht]{0.47\textwidth}
						\centering
						\includegraphics[width=0.9\textwidth]{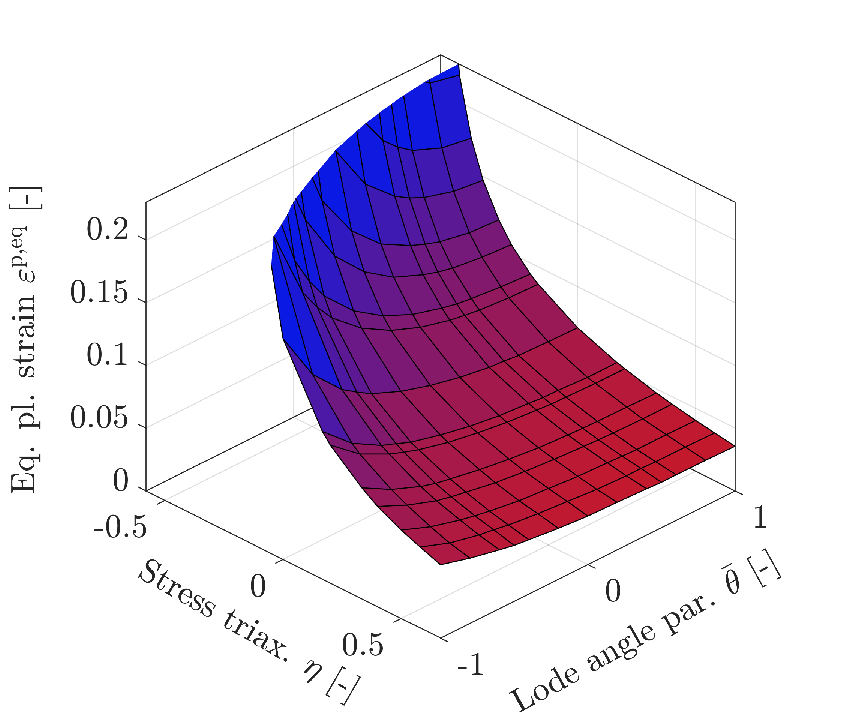}%
						\caption{\modelIa}
					\end{subfigure}
					\begin{subfigure}[ht]{0.47\textwidth}
						\centering
						\includegraphics[width=0.9\textwidth]{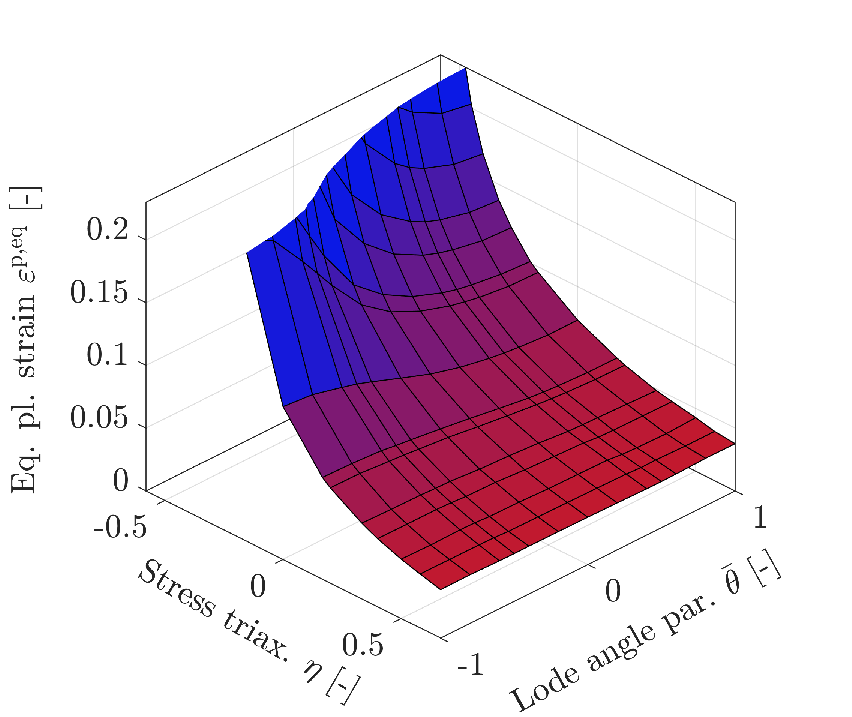}%
						\caption{\modelIb}
					\end{subfigure}
				\end{subfigure}
				\hfill
				\vspace{-30pt}
				\begin{subfigure}[ht]{\textwidth}
					\begin{subfigure}[ht]{0.47\textwidth}
						\centering
						\includegraphics[width=0.9\textwidth]{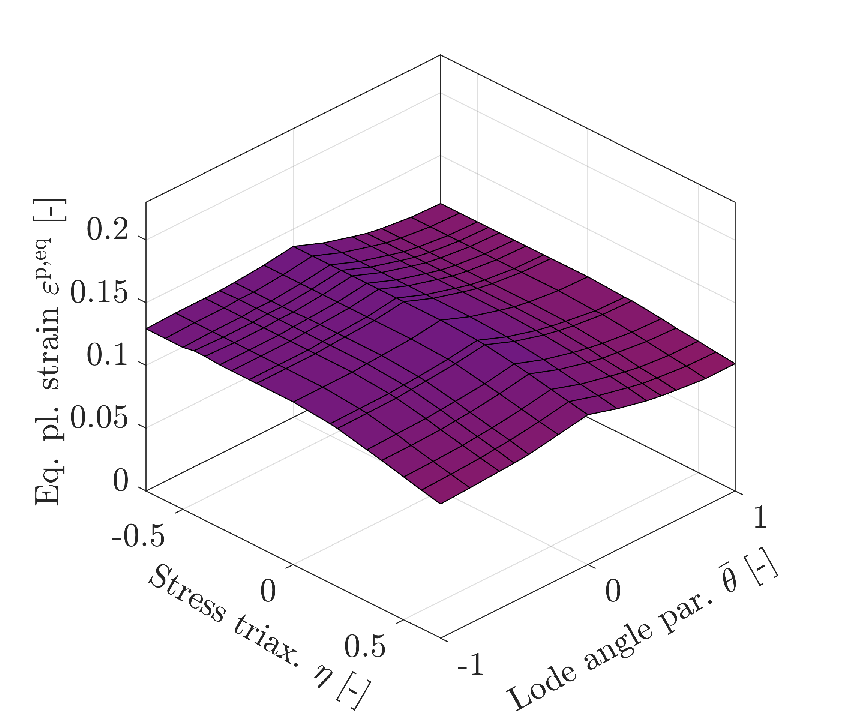}%
						\caption{\modelIIa}
					\end{subfigure}
					\begin{subfigure}[ht]{0.47\textwidth}
						\centering
						\includegraphics[width=0.9\textwidth]{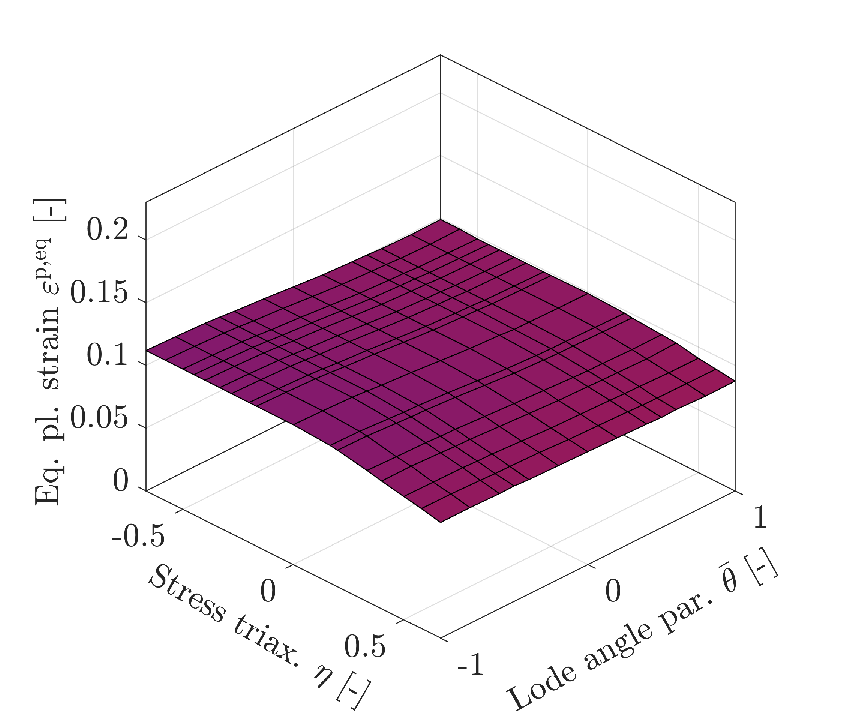}%
						\caption{\modelIIb}
					\end{subfigure}
					\makebox[0pt][r]{
						\raisebox{45pt}{%
							\includegraphics[width=0.08\textwidth]{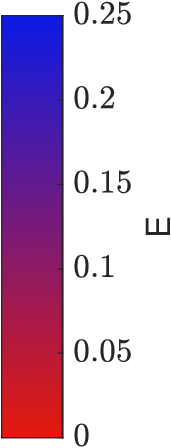}%
							\hspace{-30pt}
						}
					}%
				\end{subfigure}
				\hfill
				\vspace{0pt}
				\caption{Fracture surfaces induced by the four different CDM-based models (\modelIa, \modelIb, \modelIIa, \modelIIb) for proportional load paths. The surfaces correspond to a damage iso-surfaces of $\xi_{\mathbb{E}}=0.8$, i.e., 80$\%$ remaining directional stiffness compared to the undamaged initial state. The surfaces have been computed by prescribing the proportional load path of uniaxial tension.}
				\label{Fig:FS:FS}
			\end{figure}

			In line with \cite{wierzbicki_new_2005, BAI20081071}, the \modelI model shows the trend the smaller the stress triaxialities, the larger the equivalent plastic strains -- both for the anisotropic and for the isotropic formulation, see Fig.~\ref{Fig:FS:FS}. The \modelI model does not predict damage evolution for biaxial compression. Although the stress triaxiality is the major influencing factor of the stress state, the fracture surface corresponding to the \modelI model also depends on the Lode angle parameter with the largest sensitivity for small stress triaxialities.
			In contrast to the \modelI model, the fracture surface predicted by the \modelII model is less sensitive with respect to the stress triaxialities and the Lode angle parameter, see Figs.~\ref{Fig:FS:FS} (c) and (d). However, the Lode angle parameter again contributes separately to the damage evolution and the overall trend is similar: The smaller the stress triaxialities, the larger the equivalent plastic strains. Only for larger negative triaxialities a plateau can be seen -- in contrast to model \modelI. This plateau can be partly explained by the structure of the driving force, cf. Eq.~\eqref{eq:Ybar}. Since driving force $\bar{Y}$ is quadratic in the stress tensor, the pre-factor scaling the damage evolution is symmetric in the stress triaxiality. Clearly, the MCR effect dampens this effect.
			
			\subsubsection{Fracture surfaces for non-proportional load paths}\label{subsec:nonproportional}
				This paragraph analyzes whether the fracture surface is also meaningful for other loads paths (in contrast to the proportional ones before with constant stress triaxiality and Lode angle parameter). As a representative reference path, uniaxial tension is considered, i.e., $\B{\sigma}=\sigma_{11}\,\B{e}_1\otimes\B{e}_1$. It corresponds to a stress triaxiality of $\eta=1/3$ and Lode angle parameter of $\bar{\theta}=1$. It is compared to a second load path. The latter is inspired by uniaxial tension, the following more general stress path is considered:
				\begin{align}\label{eq:stress-strain-components-proportional2}
					\B{\sigma}(t)= \begin{bmatrix}
						\boxed{\sigma_{11}(t)} && \boxed{0} && \boxed{0}
						\\
						\text{sym}  && \boxed{\sigma_{22}(\sigma_{11}(t),\eta(t))} &&  \boxed{0}
						\\
						\text{sym} && \text{sym} && \boxed{0}
					\end{bmatrix}_{\left[\B{e}_1, \B{e}_2, \B{e}_{3} \right]}\, \text{,}
				\end{align}
				Clearly, by setting  $\eta=1/3$ and $\bar{\theta}=1$, uniaxial tension is recovered. However, the stress triaxiality is prescribed by means of a \Bezier interpolation in time. While it corresponds to uniaxial tension for the initial and the final state, i.e., $\eta(t=0)=\eta(t=1)=1/3$, it reaches its maximum deviation from this state at $t=0.5$ with a stress triaxiality of $\eta(t=0.5)=1/3-7/30=1/10$ for the first load path and $\eta(t=0.5)=1/3+7/30=17/30$ for the second load path. The Lode angle parameter is not explicitly controlled, but follows from $\sigma_{11}$ and $\eta$. Finally, it is noted that the time-scaling of $\sigma_{11}(t)$ is iteratively adapted until the state $\sigma_{11}(t=1)$ and $\eta(t=1)$ correspond to uniaxial tension and a damage threshold of $\xi_\mathbb{E} = 0.8$.
				
\begin{figure}[ht]%
					\centering
					\includegraphics[width=0.8\textwidth]{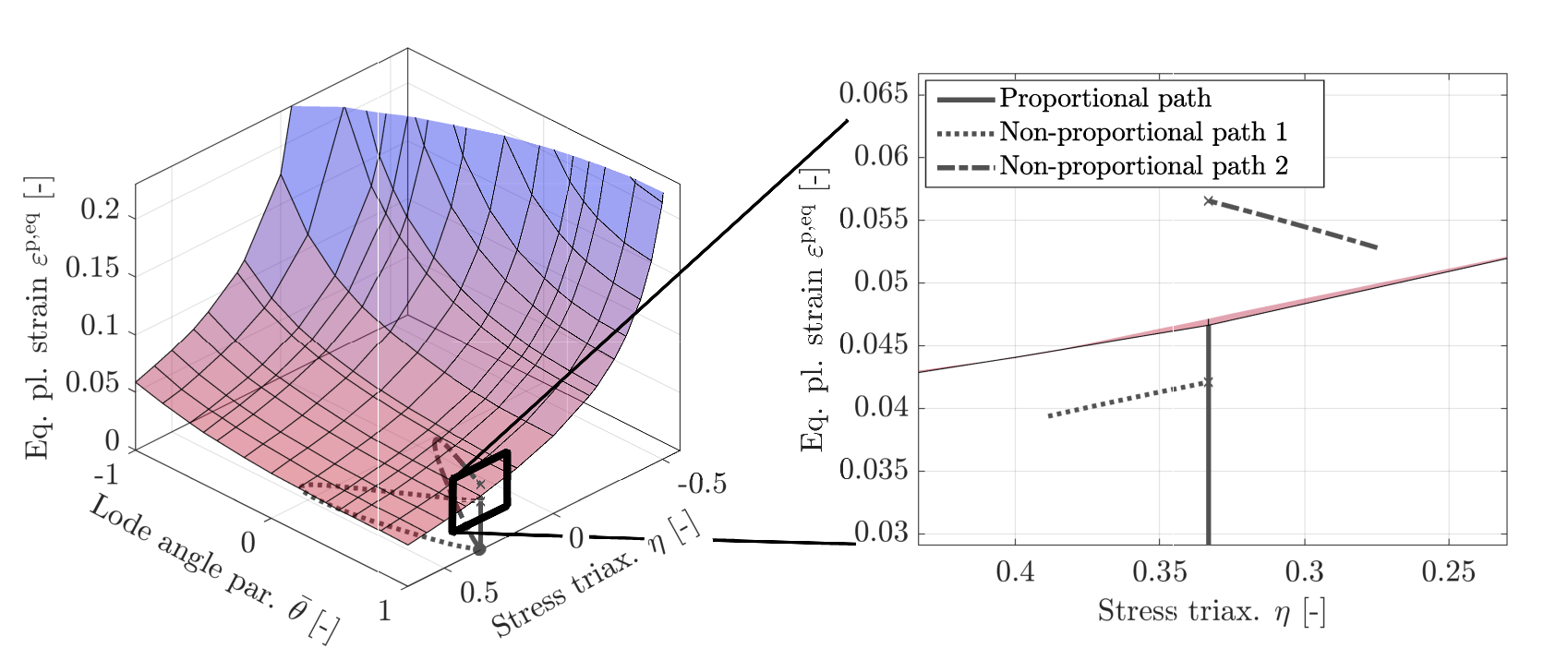}%
					\caption{Fracture surface induced by CDM-based model \modelIa. The surface is a damage iso-surfaces of $\xi_{\mathbb{E}}=0.8$, i.e., 80$\%$ remaining directional stiffness compared to the undamaged initial state. Two non-proportional load paths as defined by stress tensor~\eqref{eq:stress-strain-components-proportional} are also shown as dash-dotted and dotted lines. A zoom-in of a cut is shown on the right hand side. Accordingly, both non-proportional load paths are characterized by different equivalent plastic strains at final time $t=1$.}
					\label{Fig:FS:Damage0}
				\end{figure}

\begin{figure}[ht]%
					\centering
					\psfrag{EPSP}[c][c]{\scalebox{0.8}{Equivalent plastic strain \PEEQ [-]}}
					\psfrag{A}[c][c]{\scalebox{0.8}{\modelIa}}
					\psfrag{B}[c][c]{\scalebox{0.8}{\modelIb}}
					\psfrag{C}[c][c]{\scalebox{0.8}{\modelIIa}}
					\psfrag{D}[c][c]{\scalebox{0.8}{\modelIIb}}
					\begin{overpic}[width=0.5\textwidth]{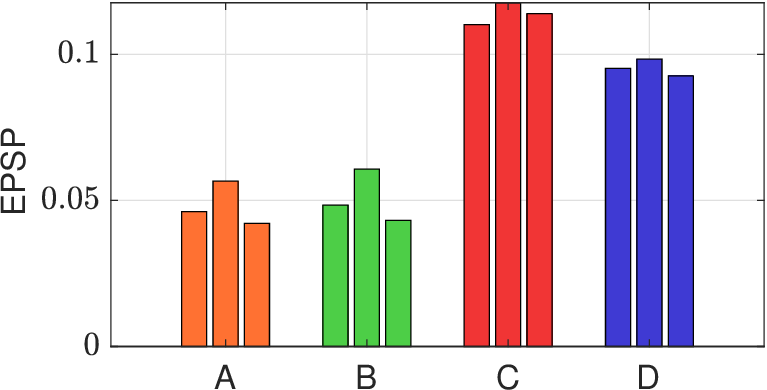}
						\put(23.5, 24){\rotatebox{90}{\scriptsize Proportional}}
						\put(28, 28){\rotatebox{90}{\scriptsize Non-prop. 1}}
						\put(32.5, 22){\rotatebox{90}{\scriptsize Non-prop. 2}}
					\end{overpic}
					\caption{Accumulated equivalent plastic strain predicted by the different CDM-based models. The strain corresponds to $\xi_{\mathbb{E}}=0.8$, i.e., iso-surface of 80$\%$ remaining directional stiffness compared to the undamaged initial state. While the left bars are associated with the proportional load path of uniaxial tension, the middle and right bars correspond to the non-proportional load paths as defined by stress tensor~\eqref{eq:stress-strain-components-proportional}.}
					\label{Fig:FS:Damage01}
				\end{figure}

				The aforementioned two different load paths are shown in Fig.~\ref{Fig:FS:Damage0} in dashed and dotted lines. They are represented by means of the triple equivalent plastic strain, stress triaxiality and Lode angle parameter. According to Fig.~\ref{Fig:FS:Damage0}, these paths are non-proportional and even more important, they lead to different equivalent plastic strains at the final time $t=1$ corresponding to $\xi_\mathbb{E} = 0.8$, see the zoom-in in Fig.~\ref{Fig:FS:Damage0} (right). The deviation of the equivalent plastic strain for these paths is above \SI{30}{\percent}. This deviation clearly confirms that a characterization and description of damage accumulation only by means of the three influencing variables -- equivalent plastic strain, stress triaxiality and Lode angle parameter -- is generally not sufficient. This holds in particular if complex, non-proportional load paths are to be analyzed. Since the path-dependent deviation of the fracture energy also relies on the underlying constitutive model, the numerical experiment has been recomputed and confirmed for all models (\modelIa, \modelIb, \modelIIa, \modelIIb); see the differences of up to \SI{30}{\percent} between the load paths in Fig.~\ref{Fig:FS:Damage01}.

				Another perspective with relevant applications in, e.g., metal forming, is the prospect of damage control. In this setting, the aim is to minimize the accumulating damage during forming. Therefore and in contrast to the previous experiment the equivalent plastic strain is prescribed (together with triaxiality and Lode angle parameter), while the final damage state is free to evolve. The time-scaling of $\sigma_{11}(t)$ is iteratively adapted to synchronize $\eta(t=1)$ with the predefined equivalent plastic strain $\bar{\ve}^\mathrm{p,eq} = 0.0462$. This specific value corresponds to the respective location on the damage iso-surface (uniaxial tension, proportional load path, \modelIa, $\eta=1/3$, $\bar{\theta} = 1$).
				
\begin{figure}[ht]%
					\centering
					\hspace{-0.6cm}
					\begin{subfigure}[h]{0.45\textwidth}
						\centering
						\includegraphics[width=\textwidth]{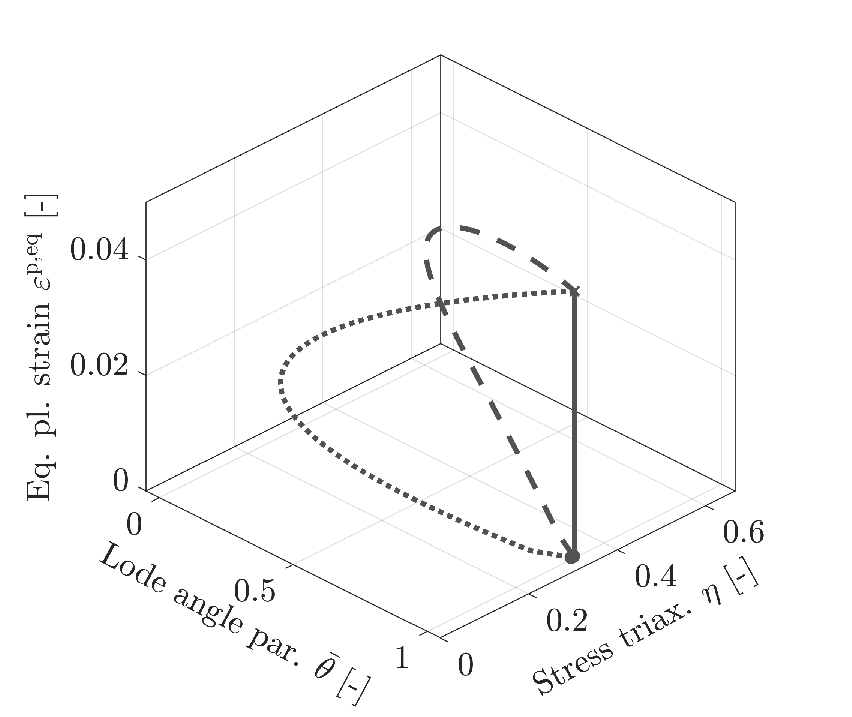}
					\end{subfigure}
					\hspace{1cm}
					\begin{subfigure}[h]{0.35\textwidth}
						\centering
						\psfrag{D}[c][c]{\scalebox{0.8}{Relative stiffness $\xi_{\mathbb{E}}$ [-]}}
						\psfrag{EP}[c][c]{\scalebox{0.8}{Equivalent plastic strain \PEEQ [-]}}
						\includegraphics[width=\textwidth]{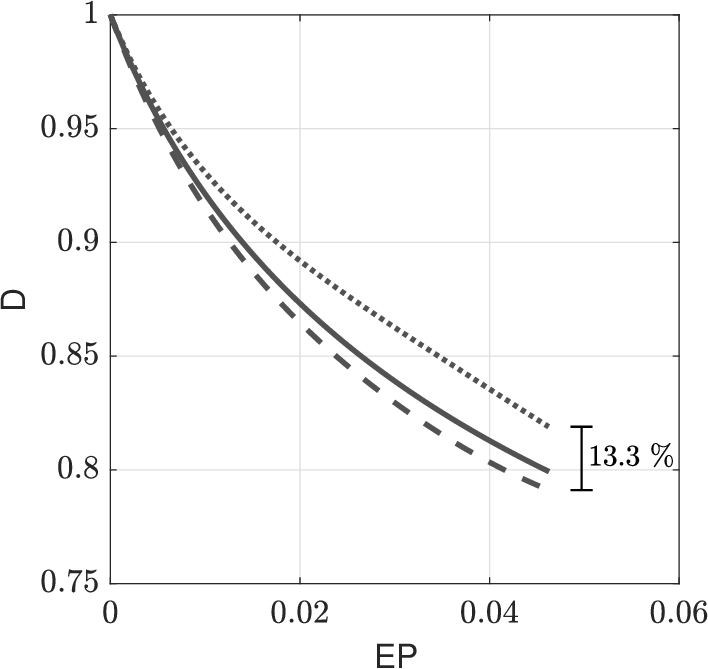}
					\end{subfigure}
					\caption{Fracture surfaces for non-proportional load paths: One proportional path (solid line) and two non-proportional paths (dashed and dotted lines) as defined by stress tensor \eqref{eq:stress-strain-components-proportional} for the \modelIa model. All paths start and end at the identical triple of stress triaxiality, Lode angle parameter and equivalent plastic strain. The normalized elastic degradation is plotted for all paths over the equivalent plastic strain.}\label{Fig:FS:triaxlodeepsp}
				\end{figure}

				The results in Fig.~\ref{Fig:FS:triaxlodeepsp} show a dependence of the damage behavior along the entire path in the triple-space of stress triaxiality, Lode angle parameter and equivalent plastic strain. Choosing the virgin material ($\xi_{\mathbb{E}} = 1$) as a reference point, lower stress triaxialities increase the remaining stiffness $\xi_{\mathbb{E}}$ by more than $9 \, \%$, while the higher stress triaxialities decrease it by $4 \, \%$.
				This viewpoint from damage control supports the previous findings of insufficient parametrization. Not only can different levels of equivalent plastic strain be achieved for a given damage threshold, but the inverse relationship holds as well. That is, different levels of damage can be achieved for a given equivalent plastic strain. This certainly confirms findings from literature and, moreover, analyzes the influence of stress triaxiality and Lode angle parameter. The developed framework now allows further quantification with load paths of 13 \% damage difference.
				The next section further constraints the framework to isolate other influences.

			\subsubsection{Non-proportional load paths with constant stress invariants}\label{subsec:nonproportional-2}
			
\begin{figure}[ht]%
					\centering
					\hfill
					\begin{subfigure}[b]{0.25\textwidth}
						\centering
						\psfrag{time}[c][c]{\scalebox{0.8}{Time $t\,[-]$}}
						\psfrag{angle}[c][c]{\scalebox{0.8}{Angle $[\text{Rad}]$}}
						\psfrag{-0.4pi}[c][c]{\scalebox{0.8}{$-0.4 \, \pi$}}
						\psfrag{-0.2pi}[c][c]{\scalebox{0.8}{$-0.2 \, \pi$}}
						\psfrag{0pi}[c][c]{\scalebox{0.8}{$0$}}
						\psfrag{0.2pi}[c][c]{\scalebox{0.8}{$0.2 \, \pi$}}
						\psfrag{0.4pi}[c][c]{\scalebox{0.8}{$0.4 \, \pi$}}
						\psfrag{alp}[l][l]{\scalebox{0.8}{$\alpha$}}
						\psfrag{bet}[l][l]{\scalebox{0.8}{$\beta$}}
						\psfrag{gam}[l][l]{\scalebox{0.8}{$\gamma$}}
						\psfrag{0}[c][c]{\scalebox{0.8}{$0$}}
						\psfrag{0.5}[c][c]{\scalebox{0.8}{$0.5$}}
						\psfrag{1}[c][c]{\scalebox{0.8}{$1$}}
						\includegraphics[width=\textwidth, height=2.8cm]{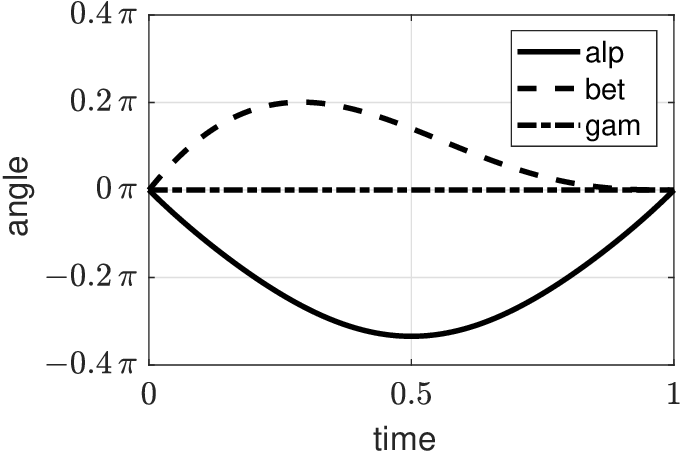}
						\vspace{-0.7cm}
						\label{Fig:IP:Angles}
					\end{subfigure}
					\hspace{0.4cm}
					\begin{subfigure}[b]{0.35\textwidth}%
						\centering
						\psfrag{RELPERCENT}[c][c]{\raisebox{10pt}{\scalebox{0.9}{\shortstack{Relative difference \\ in \PEEQ [\%]}}}}
						\psfrag{A}[c][c]{\scalebox{0.8}{\modelIa}}
						\psfrag{B}[c][c]{\scalebox{0.8}{\modelIb}}
						\psfrag{C}[c][c]{\scalebox{0.8}{\modelIIa}}
						\psfrag{D}[c][c]{\scalebox{0.8}{\modelIIb}}
						\begin{overpic}[width=0.8\textwidth]{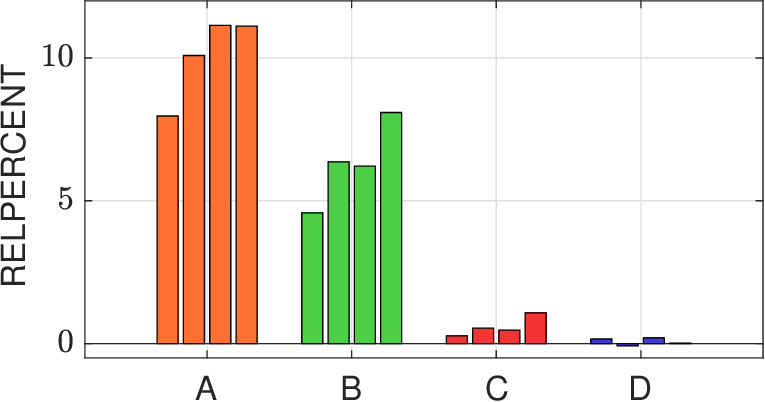}
							\put(58.5, 10){\rotatebox{90}{\scalebox{0.9}{\scriptsize Tension}}}
							\put(61.8, 11){\rotatebox{90}{\scalebox{0.9}{\scriptsize Tension+Shear}}}
							\put(65.2, 11){\rotatebox{90}{\scalebox{0.9}{\scriptsize Shear}}}
							\put(69, 13){\rotatebox{90}{\scalebox{0.9}{\scriptsize Comp.+Shear}}}
						\end{overpic}

					\end{subfigure}%
					\begin{subfigure}[b]{0.35\textwidth}%
						\centering
						\psfrag{RELPERCENT}[c][c]{\raisebox{10pt}{\scalebox{0.9}{\shortstack{Relative difference \\ in $\xi_{\mathbb{E}}$ [\%]}}}}
						\psfrag{A}[c][c]{\scalebox{0.8}{\modelIa}}
						\psfrag{B}[c][c]{\scalebox{0.8}{\modelIb}}
						\psfrag{C}[c][c]{\scalebox{0.8}{\modelIIa}}
						\psfrag{D}[c][c]{\scalebox{0.8}{\modelIIb}}
						\begin{overpic}[width=0.8\textwidth]{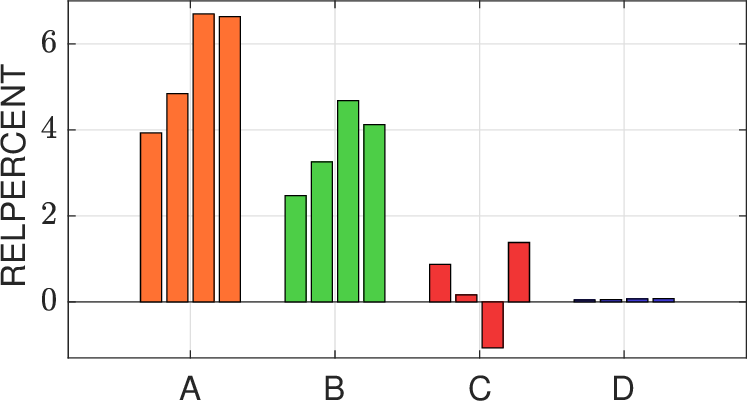}
							\put(57.75, 18.5){\rotatebox{90}{\scalebox{0.9}{\scriptsize Tension}}}
							\put(61.2, 15){\rotatebox{90}{\scalebox{0.9}{\scriptsize Tension+Shear}}}
							\put(64.6, 14){\rotatebox{90}{\scalebox{0.9}{\scriptsize Shear}}}
							\put(68, 22){\rotatebox{90}{\scalebox{0.9}{\scriptsize Comp.+Shear}}}
						\end{overpic}
						\vspace{-0.05cm}
						\label{Fig:FS:Damage3}
					\end{subfigure}
					\hfill
					\caption{Euler angles $\alpha$, $\beta$ and $\gamma$ defining a non-proportional load path according to Eq.~\eqref{Eq:spectral}. Relative difference of the equivalent plastic strain at the same final damage state $\xi_\mathbb{E} = 0.8$ (middle) and relative difference of the damage state at the same final equivalent plastic strain (right) associated with the proportional load path~\eqref{Eq:spectral-ref} and the non-proportional load path~\eqref{Eq:spectral}}
					\label{Fig:FS:Damage2}
				\end{figure}

				According to the previous section, a damage characterization by means of the triple equivalent plastic strain, stress triaxiality and Lode angle parameter is usually not sufficient for complex load paths. Since the stress triaxiality and the Lode angle parameter are already two invariants of the stress tensor one could also control all three stress invariants (instead of just $\eta$ and $\bar{\theta}$).  This is precisely the idea analyzed in this paragraph by only rotating the eigenbasis of the stress tensor. For that purpose, two different stress tensors are considered. While $\B{\sigma}$ corresponds to proportional loading, $\tilde{\B{\sigma}}$ is associated with a more complex load path. Both share the same eigenvalues $\sigma_i\of{t}=\tilde{\sigma}_i\of{t}$ (and thus the same triaxiality and Lode parameter). But they have different directions in terms of the associated eigenvectors. $\B{\sigma}$ and $\tilde{\B{\sigma}}$ can be implemented by means of the spectral decomposition, i.e.,
				\begin{align}
					\B{\sigma} &= \sum_{i=1}^{3} \sigma_i \, \B{N}_i \dyad \B{N}_i \, \text{,} \label{Eq:spectral-ref}
					\\
					\tilde{\B{\sigma}} &= \sum_{i=1}^{3} \sigma_i \, \left[\B{R} \cdot \B{N}_i\right] \dyad \left[\B{R} \cdot \B{N}_i\right] \label{Eq:spectral} \quad \text{with } \B{R}\of{\alpha, \beta, \gamma} = \B{R}_{\bar{\bar{\B{e}}}_1}\of{\alpha} \cdot \B{R}_{\bar{\B{e}}_2}\of{\beta} \cdot \B{R}_{\B{e}_3}\of{\gamma} \, \text{,} 
				\end{align}
				where $\B{N}_i$ are the eigenvectors of $\B{\sigma}$ and $\B{R}_{\B{e}_i}$  are rotation matrices depending on the $\alpha$, $\beta$ and $\gamma$. Four different mechanical tests are considered:
				%
				\begin{enumerate}
					\item Uniaxial tension: $\eta=1/3$ and $\bar{\theta}=1$
					\item Combined tension and shear: $\eta=1/6$ and $\bar{\theta}=0.506$
					\item Simple shear: $\eta=0$ and $\bar{\theta}=0$
					\item Combined compression and shear: $\eta=-1/6$ and $\bar{\theta}=-0.506$
				\end{enumerate}
				While the first eigenvalue $\sigma_1$ is increased, the second as well as the third follow from the prescribed and constant parameters $\eta$ and $\bar{\theta}$, see~\ref{Ssec:PR} for details. For the proportional load paths, stress tensor~\eqref{Eq:spectral-ref} is considered with constant eigenvectors $\B{N}_i$ and $\sigma_1$ is increased up to a damage threshold of  $\xi_\mathbb{E} = 0.8$. In the case of the non-proportional load paths, the Euler angles $\alpha$, $\beta$ and $\gamma$ evolve as illustrated in Fig.~\ref{Fig:FS:Damage2}.
				In line with the example studied in the previous paragraph, the temporal discretization of $\sigma_1(t)$ is chosen such that all Euler angles vanish at final time $t=1$ when damage threshold $\xi_\mathbb{E} = 0.8$ is reached, i.e., the time scaling of $\sigma_1$ and the Euler angles is synchronized.
				
				The bar charts in Fig.~\ref{Fig:FS:Damage2} summarize the relative differences for all four models (\modelIa, \modelIb, \modelIIa, \modelIIb) and for all four mechanical tests (uniaxial tension, simple shear, combined tension and shear, combined compression and shear). They show a relative difference of the equivalent plastic strain at final damaged state $\xi_\mathbb{E} = 0.8$ between the proportional and the non-proportional load paths. This implies again, that the chosen set of three independent stress invariants (or eigenvalues) is insufficient to fully characterize the damage evolution history. It can be seen that the anisotropic model \modelIa leads to the largest difference -- followed by its isotropic counterpart \modelIb. The sensitivity on the load path is less pronounced for the \modelII models. However, a minor influence is still evident -- even for model \modelIIb. While for some load paths, such invariant-based isotropic approximations might lead to reasonable predictions, they are generally inadequate for complex load paths.
				
				Turning the view towards damage prediction, the equivalent plastic strain will be controlled next (together with the stress triaxiality and Lode angle parameter) and the resulting damage is analyzed.
	
            	According to Fig.~\ref{Fig:FS:Damage2}, a difference in final damage measure is again observed for all load paths and damage models. The \modelIa model shows the largest deviations in the final state of over 6 \% for the paths related to shear and compression with shear. It is followed by the \modelIb and the \modelIIa models with the \modelIIb model showing the smallest load path sensitivity.
			
				Despite an identical history of all three stress invariants, the achievable equivalent plastic strain could be increased by 10 \% until the limit damage threshold. The unconsidered influences hence still bear potential for optimization of forming processes or operation conditions with respect to damage control. Stress triaxiality, Lode angle parameter and equivalent plastic strain are important control variables, but yet not sufficient. The alternative experiments showed an increase of damage measure $\xi_{\mathbb{E}}$ of up to 6 \% resulting in an improved remaining stiffness for the paths with superposed rotation.

				%

\section{Conclusion}\label{Sec:Conclusion}
	Application limits of isotropic damage characterizations for complex load paths were investigated and quantified in this paper. Since numerical experiments may depend on the underlying constitutive model, two well-established anisotropic and isotropic damage models have been calibrated and considered individually for that purpose: The \modelI model is based on the principle of strain energy equivalence and the \modelII model is the Lemaitre-model based on effective stresses. Their prediction of ductile damage has been analyzed for different load paths. 
	
	The first numerical experiment -- inspired by a coupled tension-torsion experiment -- revealed that the frequently employed recommendation "the smaller the stress triaxiality, the smaller the damage accumulation" can be wrong in general for complex load paths. In addition to the invariant stress triaxiality, the Lode angle parameter was controlled in the next example. Since the characterization of damage by means of the stress triaxiality and the Lode angle parameter is closely related to the concept of fracture surfaces, the latter have been studied in detail. Again, load paths were designed showing that a damage characterization only by means of the extended triple of influencing variables stress triaxiality, Lode angle parameter and equivalent plastic strain is often still not possible. This limitation holds notably for both damage models and even their isotropic simplifications. Deviations due to unconsidered influences on fracture (iso-)surfaces have been detected by up to 30\,\% in terms of the equivalent plastic strain and up to 13\,\% of damage, respectively.
	
	Finally, the full set of invariants of the stress tensor were controlled. This can be interpreted as an isotropic approximation. However, even in this more flexible case, load paths with different damage accumulation have been found, despite completely identical stress invariants along entire load paths. Such paths correspond to a rotation of the stress state.
	
	The numerical studies showed that damage can indeed be influenced and improved by controlling the stress triaxiality as well as the Lode angle parameter. Also the equivalent plastic strain can be chosen as a possible control variable. However, the remaining coordinates of the load paths still have great potential for further improvement of damage accumulation and thus, control. This potential is to be further studied, e.g., in the context of forming processes. Furthermore, the numerical studies are to be further confirmed by real experiments. The load paths identified in this systematic analysis are intended to guide experimental load settings for the further exploration of damage-influencing factors.


\appendix
\section{Parametrization in terms of stress triaxiality and Lode parameter}\label{Ssec:PR}
	The inversion of the equations for stress triaxiality~\eqref{Eq:stress_invariants}~and Lode parameter~\eqref{Eq:lode_angle}, i.e.,
	\begin{align}
		\sbr{\sigma_2}_{1,2} &= -\chi \, \sigma_1 \pm \sqrt{\chi^2 - \omega} \abs{\sigma_1}\\
		\sbr{\sigma_3}_{1,2} &= \sbr{a + b\,\chi} \, \sigma_1 \mp b \, \sqrt{\chi^2 - \omega} \, \abs{\sigma_1}
	\end{align}
	with 
	\begin{align*}
		a &= \frac{L+1}{L-1}, \quad b = \frac{2}{L-1}
	\end{align*}
	and
	\begin{align*}
		\chi &= \frac{18 \, \eta^2 \, \left[L-L^2-2\right] - 4\,L\,\left[L-3\right]}{18\,\eta^2 \, \left[L^2-3\right] - 2\,\left[L-3\right]^2}, \quad \omega = \frac{18\,\eta^2\,\left[L^2-L+2\right]-8\,L^2}{18\,\eta^2 \, \left[L^2-3\right] - 2\,\left[L-3\right]^2} \text{.}
	\end{align*}
allows to prescribe, and hence control, both the stress triaxiality as well as the Lode parameter. This allows to analyze load paths with identical stress triaxiality and Lode parameter with respect to damage evolution. Clearly, special care is required, if $L \rightarrow 1$.

\section*{Acknowledgements}

	Financial support from the German Research Foundation (DFG) via SFB/TR TRR 188 (project number 278868966), project C01, is gratefully acknowledged. We also gratefully acknowledge the computing time provided on the Linux HPC cluster at TU Dortmund University (LiDO3), partially funded in the course of the Large-Scale Equipment Initiative by the German Research Foundation (DFG) (project number 271512359).

\bibliography{references_main.bib}

\end{document}